\documentclass[fleqn,usenatbib]{mnras}

\usepackage{newtxtext,newtxmath}
\usepackage[T1]{fontenc}

\DeclareRobustCommand{\VAN}[3]{#2}
\let\VANthebibliography\thebibliography
\def\thebibliography{\DeclareRobustCommand{\VAN}[3]{##3}\VANthebibliography}

\usepackage{graphicx}	
\usepackage{amsmath}	
\usepackage{soul}

\newcommand{\OII}{\textsc{{\rm O}\kern 0.1em{\sc ii}}}
\newcommand{\MgII}{\textsc{{\rm Mg}\kern 0.1em{\sc ii}}}
\newcommand{\CIV}{\textsc{{\rm C}\kern 0.1em{\sc iv}}}
\newcommand{\FeII}{\textsc{{\rm Fe}\kern 0.1em{\sc ii}}}
\newcommand{\OVI}{\textsc{{\rm O}\kern 0.1em{\sc vi}}}
\newcommand{\SiIII}{\textsc{{\rm Si}\kern 0.1em{\sc iii}}}
\newcommand{\SiIV}{\textsc{{\rm Si}\kern 0.1em{\sc iv}}}

\newcommand{\ramses}{{\sc Ramses}}
\newcommand{\nocr}{no-CR}
\newcommand{\crmed}{CR$-\kappa_{\rm med}$}
\newcommand{\crhigh}{CR$-\kappa_{\rm high}$}

\usepackage{xcolor}

\title[The Effect of Cosmic Rays on the CGM]{The Effect of Cosmic Rays on the Observational Properties of the CGM}

\author[D. DeFelippis et al.]{
Daniel DeFelippis,$^{1}$\thanks{E-mail: d.defelippis@columbia.edu}
Fr\'ed\'eric Bournaud,$^{1}$
Nicolas Bouch\'e,$^{2}$
Edouard Tollet,$^{2}$
Marion Farcy,$^{3}$ 
Maxime Rey,$^{4}$ \newauthor
Joakim Rosdahl,$^{2}$ 
J\'er\'emy Blaizot$^{2}$
\\
$^{1}$Universit\'e Paris-Saclay, Universit\'e Paris Cit\'e, CEA, CNRS, AIM, 91191, Gif-sur-Yvette, France\\
$^{2}$Universit\'e Claude Bernard Lyon 1, CRAL UMR5574, ENS de Lyon, CNRS, Villeurbanne, F-69622, France\\
$^{3}$Institute for Physics, Laboratory for Galaxy Evolution and Spectral Modelling, EPFL, Observatoire de Sauverny,
Chemin Pegasi 51, 1290 Versoix, Switzerland\\
$^{4}$Department of Astronomy, Yonsei University, 50 Yonsei-ro, Seodaemun-gu, Seoul 03722, Republic of Korea\\
}

\date{Accepted XXX. Received YYY; in original form ZZZ}

\pubyear{2024}

\begin{document}
\label{firstpage}
\pagerange{\pageref{firstpage}--\pageref{lastpage}}
\maketitle

\begin{abstract}

The circumgalactic medium (CGM) contains information on the cumulative effect of galactic outflows over time, generally thought to be caused by feedback from star formation and active galactic nuclei. Observations of such outflows via absorption in CGM gas of quasar sightlines show a significant amount of cold ($\lesssim 10^4 \; \rm{K}$) gas which cosmological simulations struggle to reproduce. Here, we use the adaptive mesh refinement hydrodynamical code \ramses{} to investigate the effect of cosmic rays (CR) on the cold gas content of the CGM using three zoom realizations of a $z=1$ star-forming galaxy with supernova mechanical feedback: one with no CR feedback (referred to as \nocr), one with a medium CR diffusion coefficient $\kappa  = 10^{28} \; \rm{cm^{2}\; s^{-1}}$ (\crmed), and one with a high rate of diffusion of $\kappa = 3\times10^{29} \; \rm{cm^{2}\; s^{-1}}$ (\crhigh). We find that, for \crmed{}, the effects of CRs are largely confined to the galaxy itself as CRs do not extend far into the CGM. However, for \crhigh{}, the CGM temperature is lowered and the amount of outflowing gas is boosted. Our CR simulations fall short of the observed \MgII{} covering fraction, a tracer of gas at temperatures $\lesssim 10^4 \; \rm{K}$, but the \crhigh{} simulation is more in agreement with covering fractions of \CIV{} and \OVI{}, which trace higher temperature gas.

\end{abstract}


\section{Introduction}
\label{sec:intro}

The diffuse gas surrounding galaxies (often referred to as the circumgalactic medium, [CGM]) is made of several dynamical states (inflowing and outflow) and multiple phases \cite[e.g.][]{Tumlinson17,Faucher-Giguere23}. At any given time, gas inflowing (such as gas accretion from the cosmic web) and outflowing (such as galactic winds from supernovae and active galactic nuclei) occur in the CGM environment. Therefore, by studying the properties of a galaxy's CGM, it is possible to gain insights into these important processes of galaxy formation and evolution.

Observationally, the CGM is best studied using absorption line spectroscopy of quasar sightlines passing near foreground galaxies. From these quasar spectra, it is possible to infer column densities and the kinematics of gas along the line of sight. This technique has been developed and utilized over many decades to produce a rich collection of CGM observations from dedicated surveys like COS-Halos \citep[e.g.][]{Burchett19}, KBSS \citep[e.g.][]{Turner14}, and MEGAFLOW \citep[e.g][]{Schroetter16}. Results from these observational efforts have concluded that the CGM is composed of many different gas phases that fall into one of two broad categories. First, a cold dense phase traced by ions such as \MgII{} and \SiIII{} that has multiple kinematic components along most sightlines, suggesting a clumpy distribution within the CGM\footnote{This general characterization of the cold phase is somewhat redshift dependent: at low redshifts ($z\lesssim0.2$) not considered in this paper, the inferred densities of the cold phase are significantly lower \citep[see, e.g.][]{Werk14,McCourt18}.}, and second, a hot diffuse phase traced by ions such as \OVI{} and with broader absorption lines indicating a higher velocity dispersion and fewer spatially distinct clouds \citep[e.g.][]{Rudie19}. In order to understand the origins of and interplay between these different gas phases, it is necessary to model the CGM environment of galaxies with numerical simulations.

In recent years, much progress has been made in simulating the CGM at many different scales, ranging from idealised simulations \citep[e.g.][]{Kopenhafer23} to large cosmological simulations \citep[e.g.][]{Nelson20}. In all cases, it is necessary to model the effects of feedback from galaxies to produce a realistic CGM environment. Generally, modern simulations \cite[e.g.][]{Pillepich18a} accomplish this with feedback from two main sources: stars and active galactic nuclei (AGN). Stellar feedback usually consists of energy from supernovae explosions, as well as radiation pressure from massive stars, and is capable of launching gas out of the galaxy where it can either exit the halo completely or reaccrete onto the galaxy at a later time, producing ``fountain flows'' \cite[e.g.][]{Ubler14,DeFelippis17}. AGN feedback is usually more dominant in massive galaxies, where supermassive black holes launch fast intermittent jets from the centres of galaxies capable of drastically affecting the composition and kinematics of CGM gas over time \cite[e.g.][]{Obreja23}. With these two sources of feedback, modern cosmological simulations such as the IllutrisTNG suite \citep[][]{Marinacci18,Naiman18,Nelson18,Pillepich18,Springel18,Nelson19,Pillepich19}, EAGLE \citep{Schaye15}, and Horizon-AGN \citep{Dubois16a} are capable of producing realistic populations of galaxies in terms of quantities like stellar mass, angular momentum, and overall shape. They are also capable of generating predictions for the mass content of the CGM and outflows \citep[e.g.][]{Davies20,Mitchell20}, but these vary significantly depending on the galaxy formation model used, and are not necessarily in agreement with CGM observations.

One of the major difficulties cosmological simulations have with respect to CGM observations is related to the content of galactic outflows. Indeed, observations show that galactic outflows are multiphase, consisting of gas at high temperatures of $T > 10^{6} \; K$ \citep[e.g.][]{Chisholm18,Veilleux22} as well as low temperatures of $T \lesssim 10^4 \; K$ \citep[e.g.][]{Schroetter19,Zabl20,Avery22}. However, simulations have historically struggled to produce lower temperature ``cold'' outflows and often require outflows to be very fast and very hot in order to produce realistic galaxy populations, thus sacrificing the realism of the CGM and potentially altering the way in which the CGM and galaxy interact over longer Gyr timescales. Improvements in resolution and feedback models have reduced the gap between observed and simulated outflows \citep[e.g.][]{Nelson19,Peeples19}, but it still remains very difficult for stellar and AGN feedback alone to generate substantial and consistent cold outflowing gas.

A possible solution to this problem is to include in simulations other physically-motivated mechanisms by which gas can be expelled from the galaxy that might have been overlooked. A well-studied mechanism that has received much attention in recent years are cosmic rays (CRs) from supernovae explosions. From observations of the Milky Way, energy from CRs is expected to be in equipartition with energy from other sources like gravity and turbulence \citep{Boulares90} and to represent $\sim10\%$ of all the energy released by supernovae \citep[e.g.][]{Morlino12}, thus meaning it could significantly impact the dynamics of galaxies and the CGM. This is found to be the case: many recent studies have shown that the CGM in simulations of galaxies better reproduce absorption sightline observations from surveys like COS-Halos \cite[e.g.][]{Werk16} when CR feedback is implemented \citep[][]{Salem16,Butsky18,Ji20,Butsky22}. In these works, the CGM tends to have lower average temperatures when CR feedback is included. 

In these recent studies, CRs have been implemented in a variety of different ways. Nearly all of them centre on how to treat the CR diffusion coefficient $\kappa$, which helps set the timescale needed for the energy from CRs to escape the location they are injected in (i.e., a supernova). This diffusion can occur isotropically or anisotropically from its source, at a constant or variable rate \citep[e.g.][]{Butsky23}, and at a single energy bin or along a spectrum of possible energies \citep[e.g.][]{Hopkins21,Girichidis22}. CRs can also be transported by streaming along magnetic field lines rather than diffusion through the ambient medium \citep[e.g.][]{Wiener17}, or even a combination of both methods \citep[e.g.][]{Jiang18,Thomas19,Hopkins22}. These choices result in differing galaxy properties, particularly on the cold gas content and velocity of emerging outflows, the amount of regulation of star formation, and the gas temperature and density structure of the interstellar medium (ISM) and CGM, so constraining the possible implementations of CRs in simulations is crucial.

The numerical value of the diffusion coefficient $\kappa$ has been shown to make a huge difference on the temperature distribution and outflow rates of gas, sometimes by orders of magnitude, by setting the rate of CR transport which itself determines the shape of the CR pressure gradient. While it is possible to loosely constrain the possible values of $\kappa$ using gamma-ray luminosities from the Milky Way and local starburst galaxies \citep[e.g.][]{Chan19,Nunez-Castineyra22}, the resulting properties of the CGM are different enough that they can be used to set boundaries on $\kappa$. Following several recent analyses \citep[e.g.][]{Girichidis18,Jacob18,Dashyan20,Farcy22,Girichidis24}, we seek to study the effect of varying the diffusion coefficient $\kappa$ on the CGM by quantifying how the observable properties of the CGM, such as the covering fractions, change with $\kappa$. This will shed light on whether CR diffusion may be a key missing ingredient in galaxy formation models. 

In this paper, we study the effect CRs have on the CGM using cosmological ``zoom-in'' simulations. In particular, we study how CRs affect the CGM by comparing the covering fraction of metal lines to CGM absorption surveys such as the MEGAFLOW survey \citep{Zabl19,Schroetter21}. The structure of this paper is as follows. In Section \ref{sec:methods}, we detail the galaxy formation model and simulation setup of our analysis. In Section \ref{sec:results}, we then present results of our simulations showing the effect of CRs on the overall gas distribution in the halo (Section \ref{sec:general}), properties of the galaxy (Section \ref{sec:SFR}), properties of the CGM (Sections \ref{sec:outflows} and \ref{sec:CGM}), and CGM observables (Section \ref{sec:comparison}). In Section \ref{sec:discussion} we discuss the constraining power of CGM observations on our results and put our results in context of other recent work on CR feedback. Finally, we summarize our results and conclude in Section \ref{sec:conclusions}.

Throughout this work we assume a $\Lambda$CDM universe with dark energy fraction $\Omega_{\Lambda}=0.6825$, matter fraction $\rm \Omega_m=0.3175$, baryonic fraction $\rm \Omega_b=0.049$, Hubble constant $\rm H_0=67.11 \ km \ s^{-1} \ Mpc^{-1}$, and amplitude of density fluctuations $\sigma_8=0.83$. These parameters are consistent with results from \cite{Planck14}.

\section{Methods}
\label{sec:methods}

To study the effects of CRs on the CGM, we use cosmological zoom simulations, targeting a halo of interest and its environment with high resolution. For the simulations, we use the adaptive mesh refinement (AMR) code \ramses{} \citep{Teyssier02}. The positions of collisionless dark matter (DM) and stellar particles are evolved with a particle-mesh solver, and cloud-in-cell interpolation is used to calculate their gravitational potential. Gas evolution is computed with either an HLLC Riemann solver \citep{Toro94} for runs without CRs, or an HLLD Riemann solver \citep{Miyoshi05} for runs with CRs. The anisotropic diffusion of the CR fluid along the magnetic field is performed with the methods described by \cite{Dubois16}. To close the relation between gas internal energy and pressure, we assume an adiabatic index of $\gamma=5/3$. We initialize magnetic fields by defining a uniform grid with $1024^3$ cells and assigning random magnetic potentials to each cell interface, such that the magnetic field that arises from the curl of the potential is divergence-free. The six magnetic field components of each cell are normalized such that the initial magnetic field magnitude at a scale of 1 cMpc is $\approx10^{-17} \; \rm{G}$. We choose a relatively weak initial magnetic field to better focus on the role of CR feedback alone on the CGM. The magnetic fields are then evolved using the MUSCL scheme \citep{Teyssier06}. To identify DM haloes we use the \textsc{Adaptahop} halo finder in the most massive submaxima mode \citep{Aubert04,Tweed09}. A halo is defined as region satisfying the virial theorem that contains at least 20 DM particles and has a density 200 times the critical value.

\subsubsection*{Initial conditions and refinement scheme}

We use the \textsc{Music} package \citep{Hahn11} to generate cosmological initial conditions. \textsc{Music} allows refining of the DM mass resolution in a zoomed-in region of the simulation volume. We initially run a DM-only simulation with homogeneous resolution in a $30 \; \rm{cMpc}$$/h$ wide box to $z=0$. Then we select a target halo to be re-simulated up to $z=1$ with baryons and at a higher resolution. The criteria for our halo selection are as follows. (i) The target halo must have a $z=0$ halo mass close to $M_{\rm{target}} = 5 \times 10^{11}\ \rm M_\odot$. The target mass is chosen to simulate a halo which would likely host a galaxy with a stellar mass of $\sim10^{10} \; M_{\odot}$ at $z=1$, so as to be similar to galaxies from the MEGAFLOW survey \citep[e.g.][]{Zabl19,Schroetter21}~\footnote{In MEGAFLOW, the host galaxies associated with metal absorption lines have SFRs of 3-30 M$_\odot$~yr$^{-1}$ and $M_\star$ of 10$^{9}-10^{10.5}$ M$_\odot$. We refer the reader to Bouch\'e et al., in prep. for a more detailed presentation of the survey and observational strategy.}. (ii) The target halo must not contain any massive substructures, and (iii) it must also not contain a neighboring halo more massive than $0.2\times M_{\rm{target}}$ within three virial radii of the target halo's centre. These last two criteria are to avoid re-simulating a very complex large-scale environment at a high resolution (and high computational cost). 

\textsc{Music} is then used to progressively define a zoom region in the initial conditions with a DM particle mass of $3.5\times10^5 \; M_{\odot}$, corresponding to an effective fine resolution of $2048^3$ DM particles. This is nested inside larger regions with progressively larger DM particle masses by a factor of 8 each time, up to a coarsest particle mass of $1.4\times10^9 \; M_{\odot}$, corresponding to an effective coarse resolution of $128^3$ particles. The process of mapping out the zoomed region in the initial conditions is iterated until we confirm that the high-resolution zoom-in region has no contamination from low-resolution DM particles out to $3R_{\rm vir}$ from the centre of the targeted halo. All our production simulations use these same initial conditions and therefore model the evolution of the same galactic halo, albeit with different physics. 

The resolution of the gas and gravitational potential tracks that of the DM in the zoom-in scheme, with an effective resolution that goes from $128^3$ cells at the coarsest level, corresponding to a physical width of 350 ckpc, and progressively increasing to an effective base resolution of $2048^3$ cells in the innermost zoomed-in region, corresponding to a physical width of 22 ckpc. Within this innermost region, we also allow for adaptive refinement to a minimum cell width of 40 pc (physical, not co-moving). A cell is split into 8 equal-size children cells if $M_{\rm DM, cell} + M_{\rm b, cell}/f_b > 8 \ m_{\rm dm}$, where $M_{\rm DM, cell}$ and $M_{\rm b,cell}$ are the total DM and baryonic (gas plus stars) masses in the cell and $f_b=0.154$ is the baryon mass fraction, or if the cell width is larger than a quarter of the local Jeans length. In order to keep a roughly constant physical minimum cell width, within a factor of two, increasing maximum refinement levels are triggered with decreasing redshift. In our simulations, cell widths in the CGM generally range from $\approx1$ kpc in the inner region of the halo to $\approx3$ kpc in the outer region at the halo's virial radius. This is comparable to CGM resolutions achieved in simulations like TNG50 \citep[see Figure 1 of][]{Nelson20}, although existing simulations that focus computational efforts on the CGM itself improve resolution in the inner and outer halo by factors of $2-10$ from our values \citep[e.g.][]{Hummels19,Peeples19,Suresh19,vandeVoort19,Ramesh24}.

\subsubsection*{Thermochemistry}

We use the standard equilibrium thermochemistry of \ramses{}. Equilibrium hydrogen and helium cooling rates, via collisional ionization, collisional excitation, recombination, dielectric recombination, bremsstrahlung, and Compton cooling off the Cosmic Microwave Background, are applied using the rates listed in \cite{Rosdahl13}. For photoionization heating, we assume a \cite{Haardt96} UV background with an exponential cutoff for gas densities above $10^{-2} \ {\rm cm^{-3}}$ due to self-shielding. For $T>10^4$ K, the cooling contribution from metals is computed using tables generated with \textsc{cloudy} \citep[][version 6.02]{Ferland98}, assuming photo-ionization equilibrium with a \cite{Haardt96} UV background. For $T \le 10^4$ K, we use the fine structure cooling rates from \cite{Rosen95}, allowing the gas to cool radiatively to a density-independent temperature floor of $15$ K.  We start all our simulations with an artificially non-zero gas metallicity of $Z_{\rm init}=6.4\times 10^{-6} = 3.2\times 10^{-4} \ Z_{\odot}$ (we assume a Solar metal mass fraction of $Z_{\odot}=0.02$). This artificially non-pristine initial metallicity compensates for our lack of molecular hydrogen cooling channels in metal-free gas, allowing the gas to cool below $10^4$ K, and is calibrated so that the first stars form at redshift $z\approx 15$. 

\subsubsection*{Star formation}

Star formation is considered in cells where all the following criteria are met: the hydrogen gas density is $ > 10 \; \rm cm^{-3}$, the local overdensity is $> 200$ times the cosmic mean, the local Jeans length is smaller than one cell width, and the gas is locally convergent, and at a local maximum density. Gas is converted into stars at a rate
\begin{equation}
\dot \rho_{*} = \epsilon_* \rho / t_{\rm ff}, \label{sf_recipe.eq}
\end{equation}
where $t_{\rm ff}$ is the free-fall time and $\epsilon_*$ is the efficiency of star formation, which depends on local estimates of the gas turbulence and virial parameter \citep[for details see e.g.][]{Trebitsch17}. To follow on average the rate of star formation given by (\ref{sf_recipe.eq}), the stellar particles, each representing a stellar population, are created stochastically following a Poissonian distribution which provides the mass of the new stellar particle as an integer multiple of $m_*=400\; \rm M_{\odot}$ \citep[see][]{Rasera06}, and hence the minimum mass of a stellar particle is $m_*$. Our simulations also include runaway stars with a kick velocity of 50 $\rm{km \; s^{-1}}$, but we expect these to have little to no impact on the properties of the CGM we study in later sections \citep[see][]{Rey22}. 

\subsubsection*{Supernova feedback}

Supernova (SN) feedback is implemented with the mechanical feedback model described in \cite{Kimm14} and \cite{Kimm15}, where the SN energy is directly injected as momentum in the gas according to how well the Sedov phase is resolved. We assume four type II SN explosions per $100 \ M_{\odot}$ of stellar mass formed. This is about four times larger than predicted by the \cite{Kroupa01} initial mass function and therefore likely unrealistic, but we do this, as in the SPHINX simulations \citep{Rosdahl22} to prevent overcooling and unnaturally rapid star formation. SN explosions, each releasing $10^{51}$ ergs, are sampled in each stellar particle between 3 and 50 Myrs of its lifetime \citep{Kimm15}. Each particle returns on average 20\% of its initial mass back to the gas, with a metal yield of 7.5\%, roughly consistent with a \cite{Kroupa01} initial mass function.

\subsubsection*{Cosmic ray feedback}

CRs are modelled as a relativistic fluid that propagates anisotropically along magnetic field lines following the advection-diffusion approximation developed by \cite{Dubois16,Dubois19}, and loses energy via cooling by hadronic and coulombic interactions \citep{Gu08,Dashyan20}. This model has already been used in several works with \ramses{} \citep[e.g.][]{Dashyan20,Farcy22,Nunez-Castineyra22,Martin-Alvarez23}. The CRs are tracked as a non-thermal pressure term $P_{\rm CR}=e_{\rm CR}(\gamma_{\rm CR}-1)$, where $e_{\rm CR}$ is the CR energy density and $\gamma_{\rm CR}=4/3$ is the associated adiabatic index. The CRs are injected via each SN explosion into the gas cell hosting the exploding stellar particle, reserving 10 percent of the SN energy in each explosion to CRs.\footnote{We do not lower the CR energy injection fraction to compensate for the ``boosted'' SN feedback as this would likely render any resulting CR feedback completely inefficient. A non-boosted SN rate would result in a higher star-formation rate and potentially more cumulative CR energy injection, but measuring the size of this effect is outside the scope of this study.} We run simulations with two distinct values of the CR diffusion coefficient $\kappa=10^{28}\rm\,cm^2\,s^{-1}$ and $3\times10^{29}\rm\,cm^2\,s^{-1}$ in the simulations labelled CR-$\kappa_{\rm{med}}$ and CR-$\kappa_{\rm{high}}$, respectively. These two values are both within reasonable constraints from observations, particularly those from the Milky Way \citep[][]{Strong07,Trotta11} which generally favor a diffusion coefficient $\sim3\times10^{28} \; \rm{cm^{2} \; s^{-1}}$, and are considered to bracket regimes of slowly- and rapidly-diffusing CRs and how each of them affects the CGM \citep[see e.g.][]{Chan19,Nunez-Castineyra22}. We also run a simulation without CR feedback (called no-CR), for a comparative study of their effects. The three simulations, identical except for the inclusion of MHD and CR feedback, are listed in Table \ref{tab:simulations}. 

\begin{table}
\centering
\begin{tabular}{l|l|l} \hline \hline
Name & MHD & $\kappa$  \\ 
& & $(\rm{cm^{2} \; s^{-1}})$ \\ \hline
\nocr{} & No & -- \\ 
\crmed{} & Yes & $10^{28}$ \\ 
\crhigh{} & Yes & $3\times10^{29}$ \\ \hline \hline
\end{tabular}
\caption{Key differences between the three simulations analysed in this paper. The columns are, from left to right, (1) the name of the run, (2) whether MHD (and therefore CRs) is used in the run, and (3) the numerical value of the CR diffusion coefficient, when relevant.}
\label{tab:simulations}
\end{table}

\section{Results}
\label{sec:results}

We begin with a general description of the properties of the galaxy and CGM for the three runs (Section \ref{sec:general}). We then show more detailed results demonstrating the differences in baryonic content between the three runs, first for the stars and gas within the galaxy (Section \ref{sec:SFR}), then for gas outflowing from the galaxy (Section \ref{sec:outflows}), and finally for gas in the CGM (Section \ref{sec:CGM}). In Section \ref{sec:comparison}, we compare these simulations to CGM observations. 

\subsection{General properties}
\label{sec:general}

\begin{figure}
\includegraphics[width=\columnwidth]{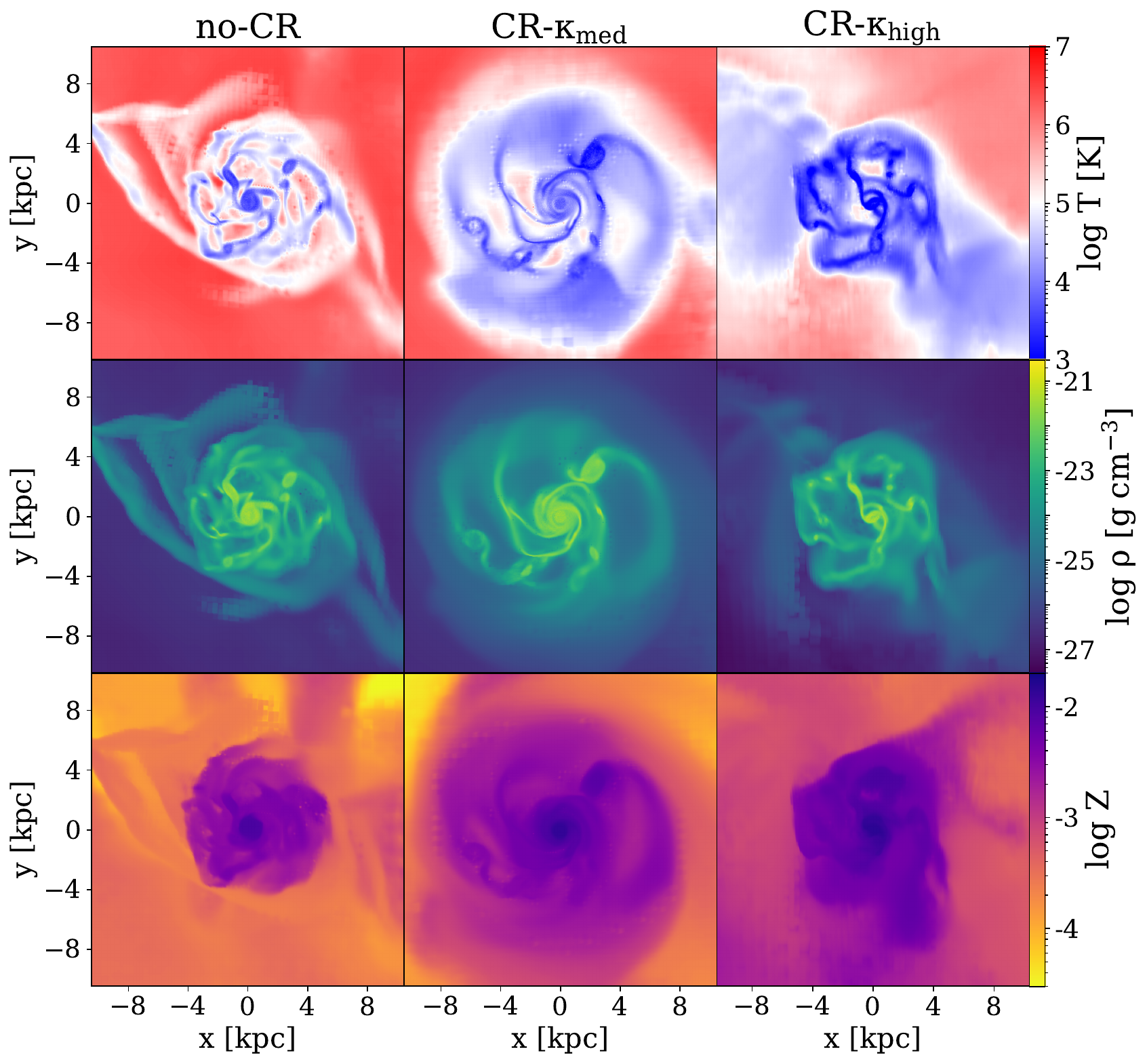} \vspace{-10pt}
\caption{Face-on projections of density-weighted temperature (top row), density (middle row), and metallicity (bottom row) for the central galaxy at $z=1$. From left to right, we show the \nocr{} run, the \crmed{} run, and the \crhigh{} run.}
\label{fig:galaxyoverview}
\end{figure}

\begin{figure}
\includegraphics[width=\columnwidth]{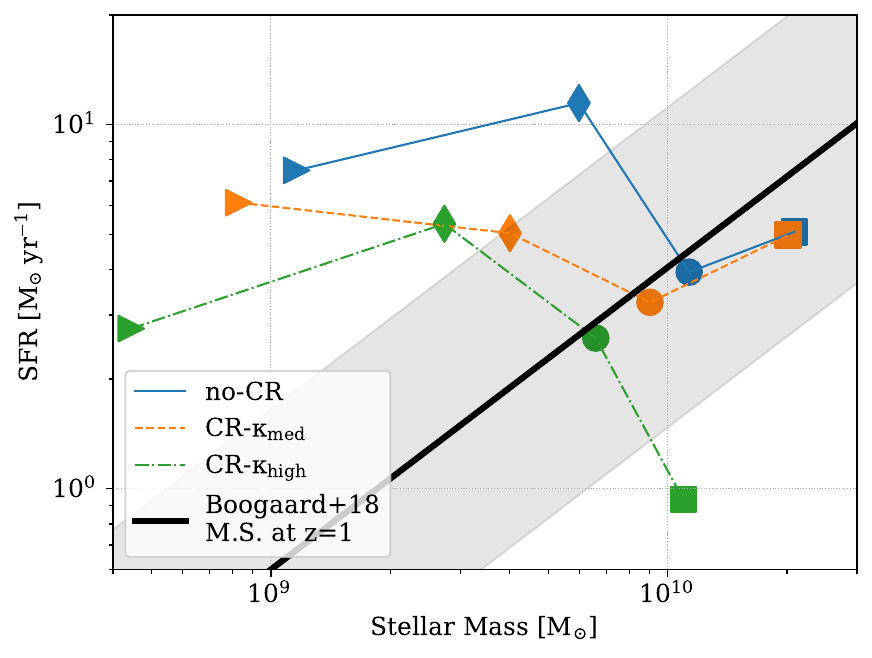} \vspace{-15pt}
\caption{Positions of the simulated galaxy, in relation to the $z=1$ star-formation Main Sequence described in \citet{Boogaard18}, at specific times for the three different runs. The symbols show locations of the runs at redshifts $z=4$ (triangles), $z=3$ (diamonds), $z=2$ (circles), and $z=1$ (squares) at the endpoint of the simulation. The SFR has been averaged for $\pm200$ Myr around each redshift to account for its high variability. According to their mass and SFR evolution, the simulated galaxies were exactly on the Main Sequence between 1.3 and 1.7 Gyr before their final ($z=1$) position.}
\label{fig:MS}
\end{figure}

\begin{figure*}
\includegraphics[width=\textwidth]{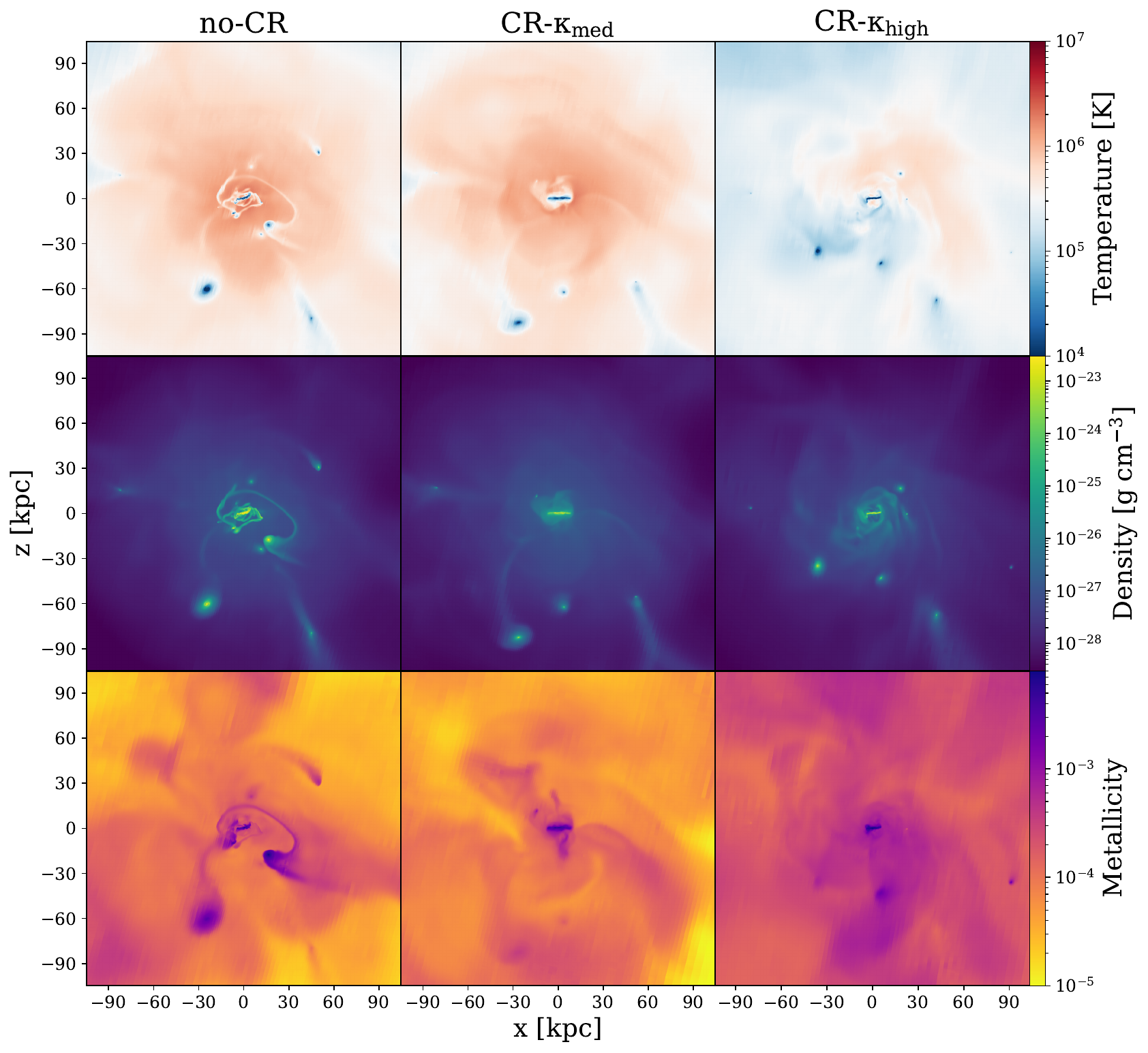} \vspace{-15pt}
\caption{Density-weighted temperature, density, and metallicity projections for the three runs at $z=1$, viewed edge-on.}
\label{fig:CGMoverview}
\end{figure*}

We start by showing face-on projections of the gas in the central galaxy in each run in Figure \ref{fig:galaxyoverview}. Qualitatively, the addition of CRs changes the appearance and extent of the central galaxy and inner CGM. With CRs, the cold ($\lesssim 10^4 \; \rm{K}$) gas is both more extended radially and distributed more smoothly within the disc, especially for \crmed{}, which also has a higher overall gas density, while the \crhigh{} gas density within the central galaxy is largely unchanged. In \crhigh{}, the central galaxy is also embedded in a somewhat higher metallicity environment compared to both other runs. Based on the morphology in Figure \ref{fig:galaxyoverview}, for this paper we define the ``galaxy'' as a cylindrical region surrounding the stellar disc with a radius $0.1 \; R_{\rm{vir}}$ and a height $0.05 \; R_{\rm{vir}}$ above and below the midplane of the disc. At $z=1$, the galaxy is a rotating gas disc $\sim10-20$ kpc across whose ISM is metal-rich and shows a substantial amount of structure. 

In Figure \ref{fig:MS}, we show the star formation rate (SFR) vs. stellar mass of the central galaxy in each run at a few selected redshifts and compare it to the observed star-formation Main Sequence of galaxies. We find that the galaxy is either slightly below (\nocr{} and \crmed{}) or significantly below (\crhigh{}) the observed $z=1$ star-formation Main Sequence at the end of the simulations. At these later times, the \nocr{} and \crmed{} runs have very similar stellar masses in their central galaxies, though the central galaxy in the \crhigh{} run is slightly less massive. We examine these differences in more detail in Section \ref{sec:SFR}. 

In Figure \ref{fig:CGMoverview}, we show density-weighted edge-on projections of the main halo in each of the three runs out to its virial radius, which is $\approx100$ kpc. In all three runs, we see the cold ($T \leq 10^4 \; K$), dense, and metal-rich galaxy in the centre which is clearly distinct from the warmer, more diffuse, and lower metallicity CGM. However, closer inspection reveals differences between the three runs. In both the \nocr{} and \crmed{} runs the CGM is dominated by relatively diffuse gas with a mean temperature of around $10^6 \; K$ apart from cold gas-dominated satellites. The typical temperature in the halo of the \crhigh{} run though is nearly 1 dex lower. This indicates that the mere addition of CRs in \crmed{} is not enough to alter the phase of the CGM: a minimum level of diffusivity must be necessary for the CRs to be able to escape from the galaxy and influence the temperature of the surrounding medium. The gas density largely follows the temperature projections: the densest gas within the galaxy and bound to satellites is also the coldest. Unlike the temperature, the gas density shows little variation between the three runs at any location in the halo, although the density distribution in the outer halo of the \crhigh{} run appears slightly less smooth than it does in the other two runs. Finally, we examine the metallicity of CGM gas in the bottom panels. Here, as in the temperature projections, we find that metallicity distributions of the \nocr{} and \crmed{} runs are very similar, but the \crhigh{} run shows a dramatically higher metallicity throughout the entire CGM. 

\subsection{Star formation}
\label{sec:SFR}

\begin{figure}
\includegraphics[width=\columnwidth]{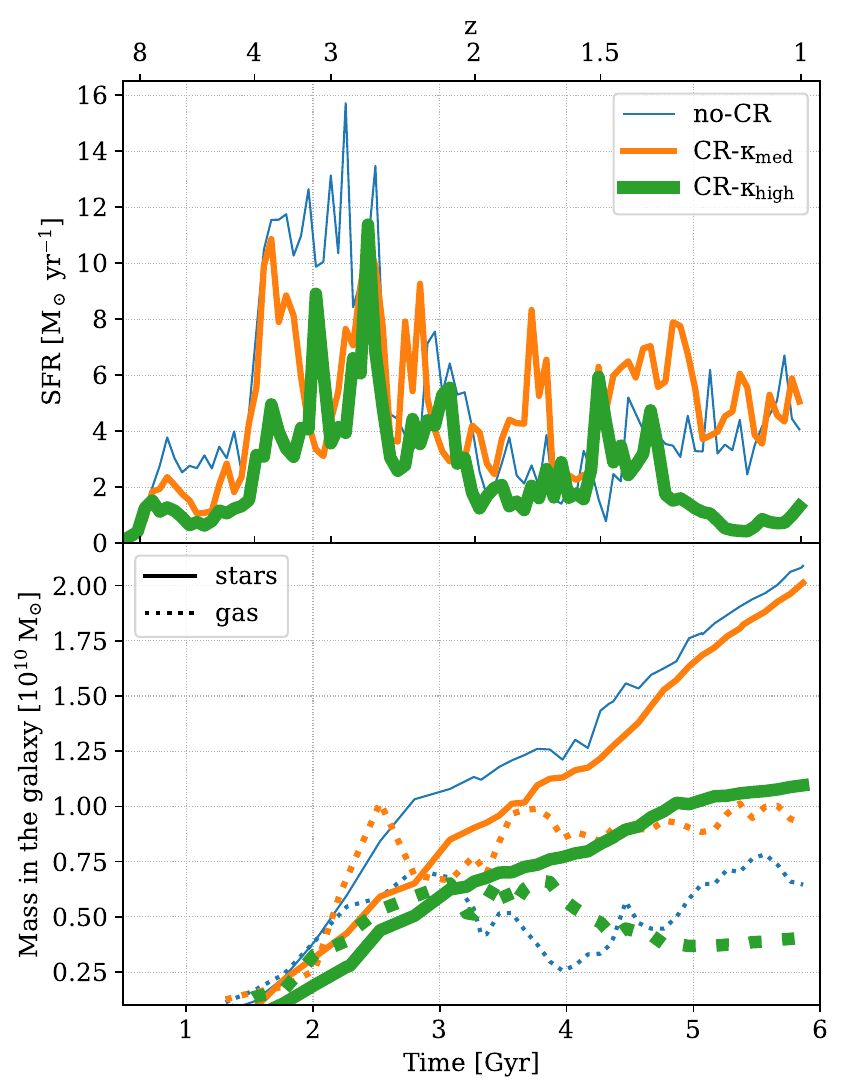} \vspace{-15pt}
\caption{Top: SFRs of the central galaxy (defined as a cylindrical region surrounding the stellar disc with a radius $0.1 \; R_{\rm{vir}}$ and a height $0.05 \; R_{\rm{vir}}$ above and below the midplane of the disc) as a function of time for the \nocr{} (blue), \crmed{} (orange) and \crhigh{} (green) runs. CRs reduce the SFR most significantly in the first 2.5 Gyr of the runs. Bottom: stellar and gas mass in the central galaxy as a function of time for the same three runs.}
\label{fig:SFR_stellarmass}
\end{figure}

We start our more detailed investigation of the effects of CRs on the galaxy by examining their effect on the SFR over the entire length of each run, which we show in the top panel of Figure \ref{fig:SFR_stellarmass}. The \nocr{} run is characterized by a SFR that varies between $\sim2-8 \; M_{\odot} \; \rm{yr^{-1}}$ for most of its history, except for a $\approx1 \; \rm{Gyr}$ time period around $z\approx 3$ where the SFR jumps above $10 \; M_{\odot} \; \rm{yr^{-1}}$. The galaxy in this run is therefore unambiguously star-forming with a very bursty star-formation history. In the \crmed{} run, the addition of CRs lowers the SFR at early times in the simulation, especially during the ``starburst'' period around $z\approx3$, but otherwise maintains the typical value and burstiness at later times. The \crhigh{} run behaves almost the same as the \crmed{} run, though the typical SFR after the starburst period is lower and drops down below $2 \; M_{\odot} \; \rm{yr^{-1}}$ at the end of the simulation. This behavior is generally consistent with the effect of CRs found in previous works such as \cite{Hopkins20}, who find that higher values of CR diffusion more effectively suppress the SFR of MW-mass galaxies.

In the lower panel of Figure \ref{fig:SFR_stellarmass}, we plot the stellar and gas masses of the central galaxy in each run over time, which shows the cumulative effect of the star formation and accretion histories. As reflected in the SFRs, the \nocr{} run has the strongest period of growth from $z\approx 4-3$ before settling down slightly, while the two CR runs' stellar masses grow more steadily throughout the simulation and are consistently below the level of the \nocr{} run. By the end of the simulations, \nocr{} has a factor of $\sim2$ higher stellar mass than \crhigh{}, whereas \crmed{} is only slightly less massive than \nocr{}. This factor of 2 difference appears to develop during the period of high star formation, and then it remains relatively constant afterwards. The gas mass in all three runs reaches a peak after $\approx3$ Gyr and then either fluctuates around that value as in the \nocr{} and \crmed{} runs or slowly decreases with time as the \crhigh{} run does.

\subsection{Outflows}
\label{sec:outflows}

\begin{figure} 
\includegraphics[width=\columnwidth]{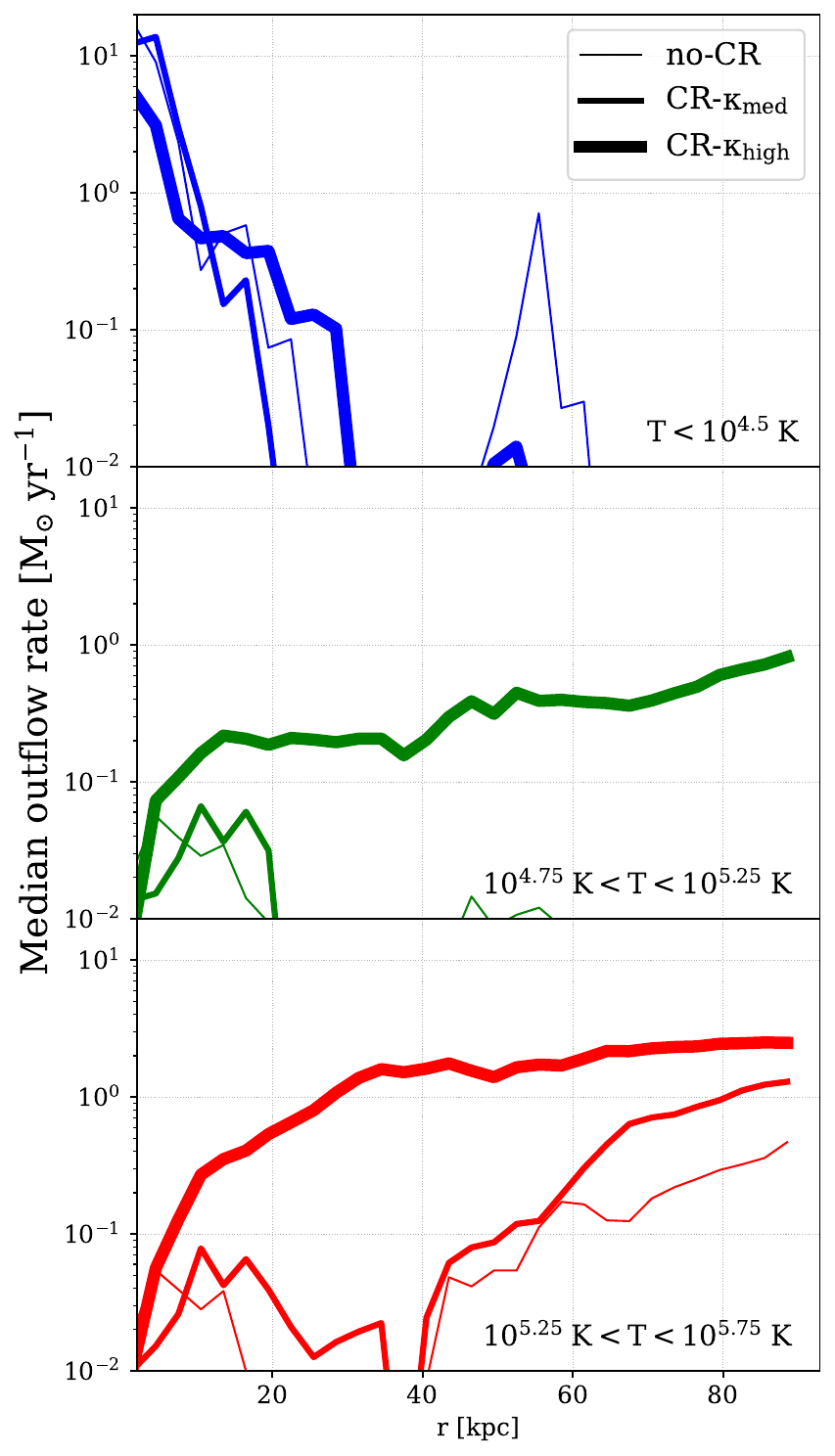} \vspace{-15pt}
\caption{Median outflow rates of five snapshots between redshifts $1 < z < 1.1$ of gas in spherical shells surrounding the central galaxy for the no-CR (thin lines), CR-$\kappa_{\rm{med}}$ (medium lines) and CR-$\kappa_{\rm{high}}$ (thick lines) runs. The gas is separated by temperature ranges roughly corresponding to gas observed in \MgII{} (top panel), \CIV{} (middle panel), and \OVI{} (bottom panel).}
\label{fig:outflow_rates}
\end{figure}

In this section, we examine the properties of outflowing gas and attempt to connect galactic outflows to star formation in the galaxy. We calculate the median outflow rate over the final five snapshots of the runs, representing a narrow redshift range of $1 < z < 1.1$. This is a large enough number of snapshots such that transitory features of the gas distribution (e.g. a short-lived tidal tail) will be removed, and a small enough number to also ensure that we do not include cosmological evolution in the median. In Figure \ref{fig:outflow_rates}, we show these median outflow rates (i.e. gas with a positive radial velocity) in radial bins around the central galaxy for the three runs, separated by temperature ranges that roughly correspond to commonly observed ions (\MgII, \CIV, and \OVI). For the coldest gas, the three runs behave very similarly overall, with high outflow rates of $\sim10 \; M_{\odot} \; \rm{yr}^{-1}$ very close to the centre of the galaxy which quickly drop to below $0.1 \; M_{\odot} \; \rm{yr}^{-1}$ by 20 kpc. However, the slope of this outflow rate is noticeably shallower for the \crhigh{} run, resulting in values smaller than the other two runs within $\approx 10$ kpc and larger than the other two runs at the very inner edge of the CGM. Above this radius, there is no appreciable outward-moving cold gas in any of the runs, except for that in a satellite galaxy with an overall positive radial velocity in the \nocr{} run. In the lower two panels, however, we see much more significant differences in the \crhigh{} run compared to the other two runs. For ``warm'' gas, which has no substantial outflowing mass anywhere in the CGM for the \nocr{} and \crmed{} runs, the \crhigh{} run shows an outflow rate of $0.2-0.9 \; M_{\odot} \; \rm{yr^{-1}}$ increasing with radius in the CGM. This is also seen in hotter gas, where the outflow rate in \crhigh{} reaches and maintains order unity by $\approx30$ kpc whereas both of the other runs remain below \crhigh{} and only approach it near the virial radius. 

We find (but do not show) that the gas velocities contributing to these outflow rates are typically small with median values up to 20 $\rm{km \; s^{-1}}$. In the \crhigh{} run, the amount of mass moving at all positive radial velocities is larger, including some material moving above 100 $\rm{km \; s^{-1}}$, at all radii in the halo, meaning that the larger outflow rates come from both more outflowing mass and faster outflowing mass. These strong CR-driven outflows in the \crhigh{} run also highlight the fact that its central galaxy's steadily declining gas mass in Figure \ref{fig:SFR_stellarmass} is due to gas expulsion via CR feedback rather than consumption by star formation. We note however that all radial velocity distributions have a negative median value, indicating that slow, steady accretion remains the dominant process occurring in the halo, even when CR feedback is operating.

\subsection{CGM properties}
\label{sec:CGM}

\begin{figure*}
\includegraphics[width=\textwidth]{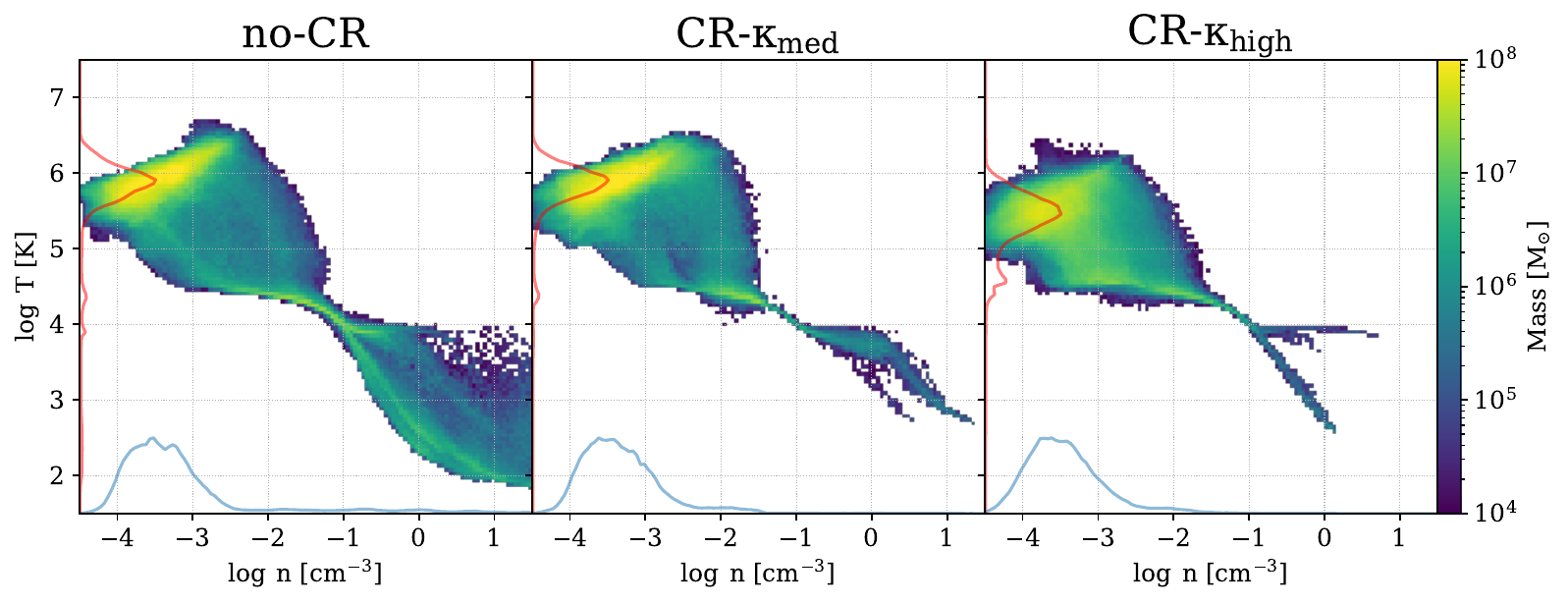} \vspace{-15pt}
\caption{Median temperature-number density phase diagrams of the central galaxies' CGM for the \nocr, \crmed, and \crhigh{} runs for the five snapshots between redshifts $1 < z < 1.1$, where the ISM (i.e. all gas within the cylindrical region of the galaxy) has been removed. The colour shows the gas mass distribution. The red and blue curves on each panel are normalized probability density functions of the temperature (red) and density (blue) of the CGM gas. CRs almost completely remove gas below $10^4 \; \rm{K}$ from the CGM, but only $\kappa_{\rm{high}}$ noticeably changes the phase structure of the diffuse CGM.}
\label{fig:phasediagrams}
\end{figure*}

We now turn to the CGM itself and highlight similarities and differences between the three runs. In Figure \ref{fig:phasediagrams}, we show stacked mass-weighted temperature-number density phase diagrams for the CGM of the central galaxy in the three runs for the same narrow redshift range as Figure \ref{fig:outflow_rates}. In these phase diagrams, we have removed gas that is in the central galaxy as defined in Section \ref{sec:general} so as to only consider CGM gas. First, we see that the \nocr{} run contains a substantial amount of gas with $T < 10^4 \; K$ in the CGM whereas the two CR runs contain much less gas below this temperature. We see from the maps in Figure \ref{fig:CGMoverview} that this is due to a combination of the dense tidal tails surrounding and directly connected to the galaxy and the more massive and numerous satellites found in the \nocr{} run. The same structures are significantly reduced in number and density for both of the CR runs. Thus, the CRs are likely able to help dissipate the very dense and cold gas found primarily in satellites and some tidal tails. Most of the CGM gas in all of the runs, however, is $> 10^4 \; K$. For the \nocr{} and \crmed{} runs, the phase structure of this hotter diffuse gas is nearly identical. The peak of these temperature distributions is $\approx 10^6 \; K$, and the diffuse CGM component spans roughly 1.5 orders of magnitude in density. In the \crhigh{} run however, the diffuse CGM has a noticeably cooler average temperature: it is peaked at a lower temperature of $\approx 10^{5.5} \; K$ and the temperature distribution is noticeably wider, resulting in a more substantial amount of gas at temperatures between $10^4$ and $10^5 \; K$ than the other two runs have. This behavior is qualitatively similar to other recent studies of CRs, which generally find that the CGM is cooler when CR feedback is included \citep[e.g.][]{Ji20,Farcy22}.

\begin{figure}
\includegraphics[width=\columnwidth]{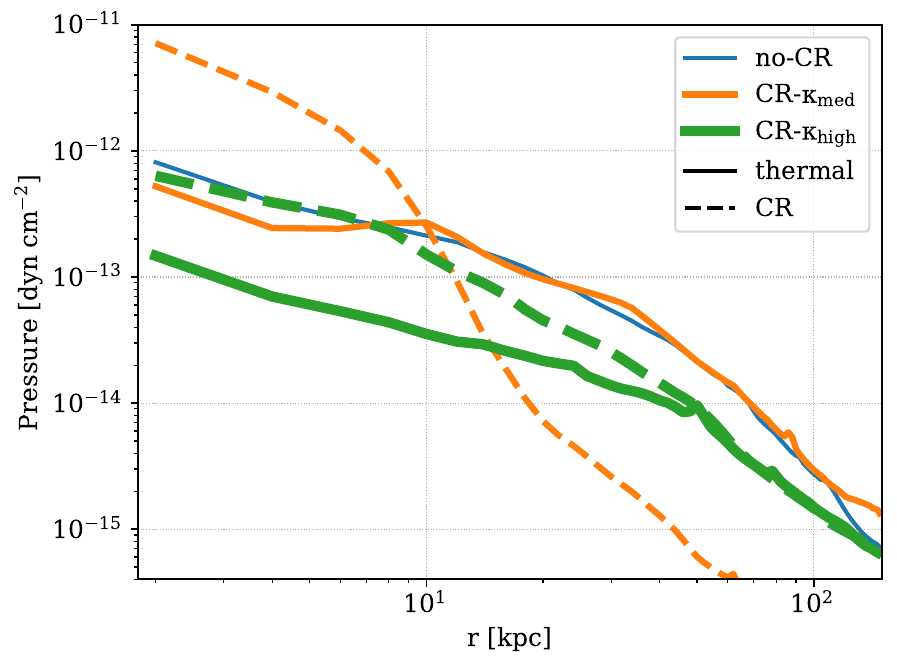} \vspace{-15pt}
\caption{Mean gas pressure vs. radius for the three runs at $z=1$. The \nocr{} run only has thermal pressure (solid lines) but the two runs with CRs additionally have non-thermal pressure (dashed lines).} 
\label{fig:pressure}
\end{figure}

We can understand the varying effects of the CR diffusion by examining the mean pressure profiles of CGM gas. As the initial magnetic fields in these simulations are very weak, the total pressure profiles should all be roughly equal due to hydrostatic pressure equilibrium. We indeed find this to be true outside of the galaxy for $r > 10$ kpc, but at smaller radius the \crmed{} run has a higher total pressure by nearly an order of magnitude. To see why this is, we plot mean pressure profiles (the median profiles are almost identical) separated into thermal and non-thermal components in Figure \ref{fig:pressure}. Compared to the \nocr{} run, the \crmed{} run has a thermal pressure profile that is almost identical, likely because the similar cumulative star formation between the two runs released a similar amount of energy from supernovae. However, the \crmed{} run's non-thermal pressure exceeds the thermal pressure within the galaxy by an order of magnitude, and is the source of the discrepancy in the total pressure profiles. While it is unrealistic for such a high non-thermal pressure to persist in the galaxy without first losing energy, it does not affect the CGM at all: beyond $\approx15$ kpc, the thermal pressure is completely dominant, as would be expected for CRs that remain largely trapped within the galaxy due to lower diffusivity. In the \crhigh{} run, however, the non-thermal pressure is the dominant source of pressure in the galaxy and the inner $\approx40$ kpc of the CGM. At larger radii, the thermal and CR pressures are comparable, thus allowing slightly colder gas not heated up from the surrounding thermal pressure to exist in the entire CGM and boosting the amount of outflowing gas seen in ions like \CIV{} and \OVI{} as shown in Figure \ref{fig:outflow_rates}. This also explains why the \crmed{} run's CGM has the same temperature as the \nocr{} run's CGM: trapped CRs only affect the properties of the ISM and largely leave the CGM unaffected.

\subsection{Comparison to observations}
\label{sec:comparison}

Having provided a description of the effects CRs have on our simulated galaxy and its CGM, we now seek to compare the CGM covering fractions to those observed in quasar absorption line studies, such as the MUSE GAs FLOw and Wind (MEGAFLOW) survey~\footnote{This survey was designed to study the CGM properties around star-forming galaxies using the Multi Unit Spectroscopic Explorer (MUSE, \citealt{Bacon10}) spectrograph on the Very Large Telescope (VLT) towards two dozen quasar sight-lines (Bouch\'e et al. in prep.).} for \MgII{} (Bouch\'e et al. in prep.). In particular, \citet{Schroetter21} investigated the \MgII{} (and \CIV{}) covering fraction of star-forming galaxies at $1<z<1.4$ using  $\sim100$ \ion{Mg}{II} absorption lines (rest equivalent width $W_r^{\lambda2796} > 0.5 - 0.8$ \AA) and $\sim 200$ star-forming galaxies within 250 kpc of the quasar sight-lines. In addition, we also use covering fractions for the higher ions \OVI{} from \citet{Kacprzak15} and \citet{Tchernyshyov23}, whose host galaxies have a similar $M_\star$ of $10^{10}-10^{11}$ M$_\odot$ and $0.1\lesssim z \lesssim0.7$, and \CIV{} from \citet{Bordoloi14}, whose host galaxies have $M_\star$ of $10^{8.5}-10^{10}$ M$_\odot$ and $z < 0.1$.

\begin{figure*}
\includegraphics[width=\textwidth]{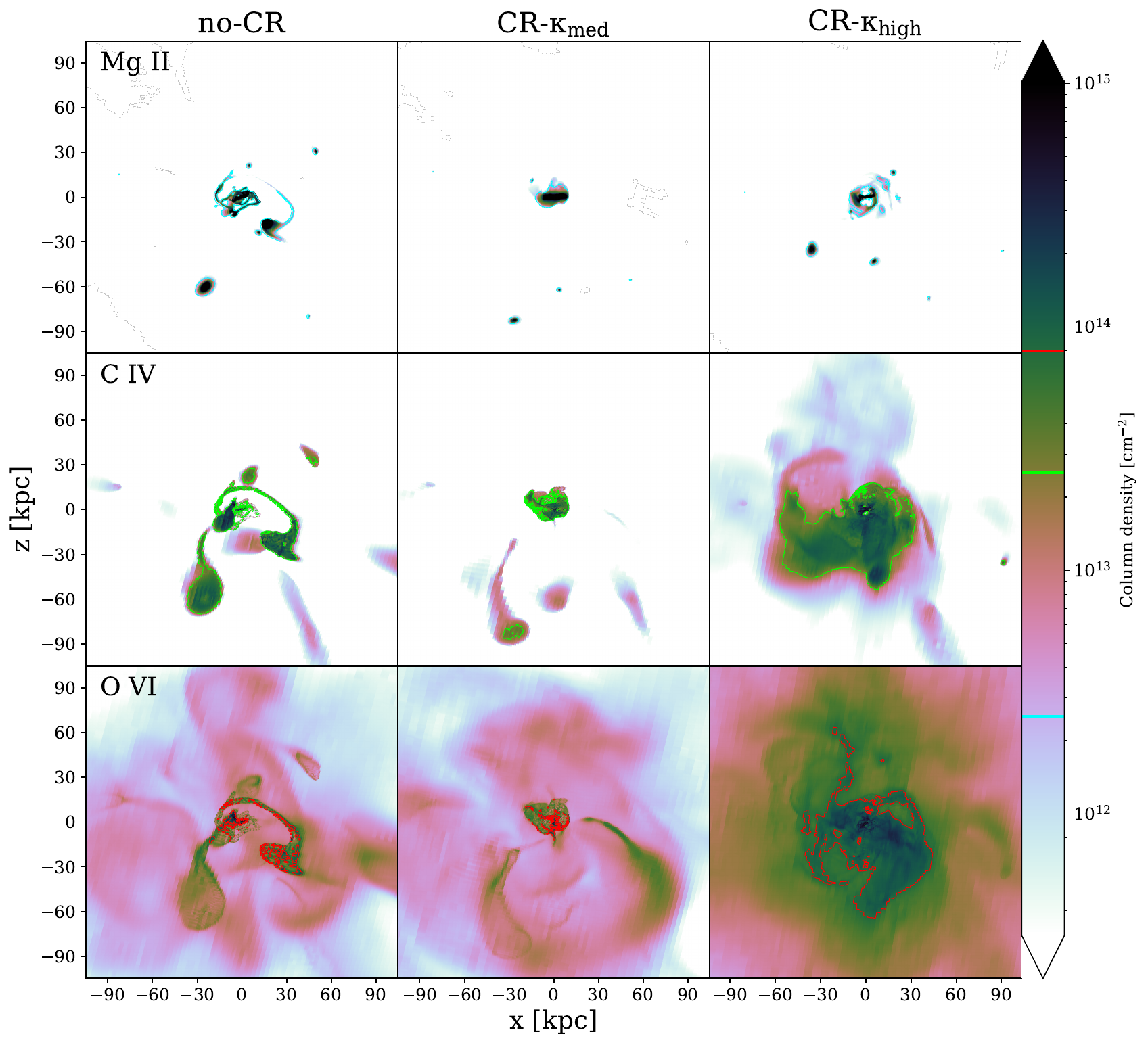} \vspace{-15pt}
\caption{Column density maps of the central haloes in the \nocr{} (left), \crmed{} (middle), and \crhigh{} (right) runs at $z=1$, viewed edge-on. Rows show \MgII{} (top), \CIV{} (middle), and \OVI{} (bottom). Blue, green, and red contours highlight the minimum absorber column densities observed in recent surveys from MEGAFLOW \protect\citep{Schroetter21} and \protect\cite{Kacprzak15}. In all runs, nearly all \MgII{} is concentrated within the galaxy, satellites, or in tidal tails, and is rare in the CGM. \CIV{} is slightly more extended in the CGM, especially in the \crhigh{} run, and \OVI{} is more volume-filling in all runs, but most significantly in \crhigh{}.} 
\label{fig:Nmaps}
\end{figure*}

In order to compare our simulations to these quasar sightline observations, we use \textsc{Trident} \citep{Hummels17} to populate our simulations with specific ions using parameters derived from \textsc{Cloudy} \citep{Ferland13} ionization tables. For this paper, we focus on three ions commonly observed in absorption -- \MgII, \CIV, and \OVI{} -- because they each trace a different temperature phase of the gas ($\lesssim 10^4 \; K, \; \approx 10^4-10^5 \; K$, and $\approx 10^{5.5} \; K$ respectively, from \citealp{Tumlinson17}). In Figure \ref{fig:Nmaps}, we show column density maps of these three ions for the three runs at $z=1$. We show the observational column density cutoffs for the three ions (\citealt{Schroetter21} for \MgII{} and \CIV{} and \citealt{Kacprzak15} for \OVI) as coloured contours. First, we see that at distances $>25$ kpc from the galaxy all three runs exhibit a CGM similarly devoid of \MgII{} except for the presence of satellites. This indicates that in addition to not changing the phase structure, neither run with CRs is any more effective than the default feedback model at pushing \MgII{} gas out of the galaxy to large distances in the CGM. Within $<25$ kpc from the galaxy though, all three runs have different \MgII{} properties. In \nocr, \MgII{} traces the clear tidal tails seen in Figure \ref{fig:CGMoverview} whereas in \crmed{} those tidal tails are not cold or dense enough to absorb \MgII, and there is a sudden column density drop off at the edge of the disc. In \crhigh, small \MgII{} column densities extend slightly beyond the disc but they merely approach and do not exceed current observational column density limits from MEGAFLOW. 

Next, we examine the \CIV{} distributions around the galaxy (middle row of Figure \ref{fig:Nmaps}). For the \nocr{} run, most of the highest \CIV{} column densities overlap with where the \MgII{} is (i.e. in satellites and tidal tails) but it extends beyond where the \MgII{} stops, indicating the cold structures are immediately surrounded by a warmer interface. This warmer and more diffuse gas is also being stripped from satellites further out in the CGM. In the \crmed{} run, we again see an abrupt drop off at the edge of the disc similar to what is seen in \nocr{} and all of the \MgII{} maps, as well as a diffuse envelope being stripped from satellites in the same way. However, the \crhigh{} run shows a drastically different distribution of \CIV{} with higher column densities ($\gtrsim 10^{13} \; \rm{cm^{-2}}$) out to 50 kpc and lower values ($\lesssim 10^{12} \; \rm{cm^{-2}}$) that reach the virial radius and are close to volume-filling. Satellite galaxies do not stand out in this \CIV{} map in the same way they do in \nocr{} and \crmed, indicating that this phase of gas is found more in the diffuse ``smooth'' component of the CGM (as suggested by Figure \ref{fig:phasediagrams}) and is not merely a warm ``interface'' between cold $T \lesssim 10^4 \; K$ structures and hot $T > 10^{6} \; K$ gas. Furthermore, the highest \CIV{} column densities in this panel also trace regions of high metallicity seen in Figure \ref{fig:CGMoverview}.

Finally in the bottom row of Figure \ref{fig:Nmaps}, we show \OVI{} column densities. In all three runs, this phase of gas is volume-filling, although we again see a strong dichotomy between the \crhigh{} run and the other two runs. In the former, \OVI{} picks up high-metallicity gas at temperatures $< 10^6 \; K$ that is distributed throughout the halo as seen in Figure \ref{fig:CGMoverview}, while in both of the latter, the highest \OVI{} column densities primarily overlap with satellites and tidal tails as is the case with both \MgII{} and \CIV{}, and at all other locations in the halo \OVI{} has at least 1 dex smaller column densities. 

\begin{figure}
\includegraphics[width=\columnwidth]{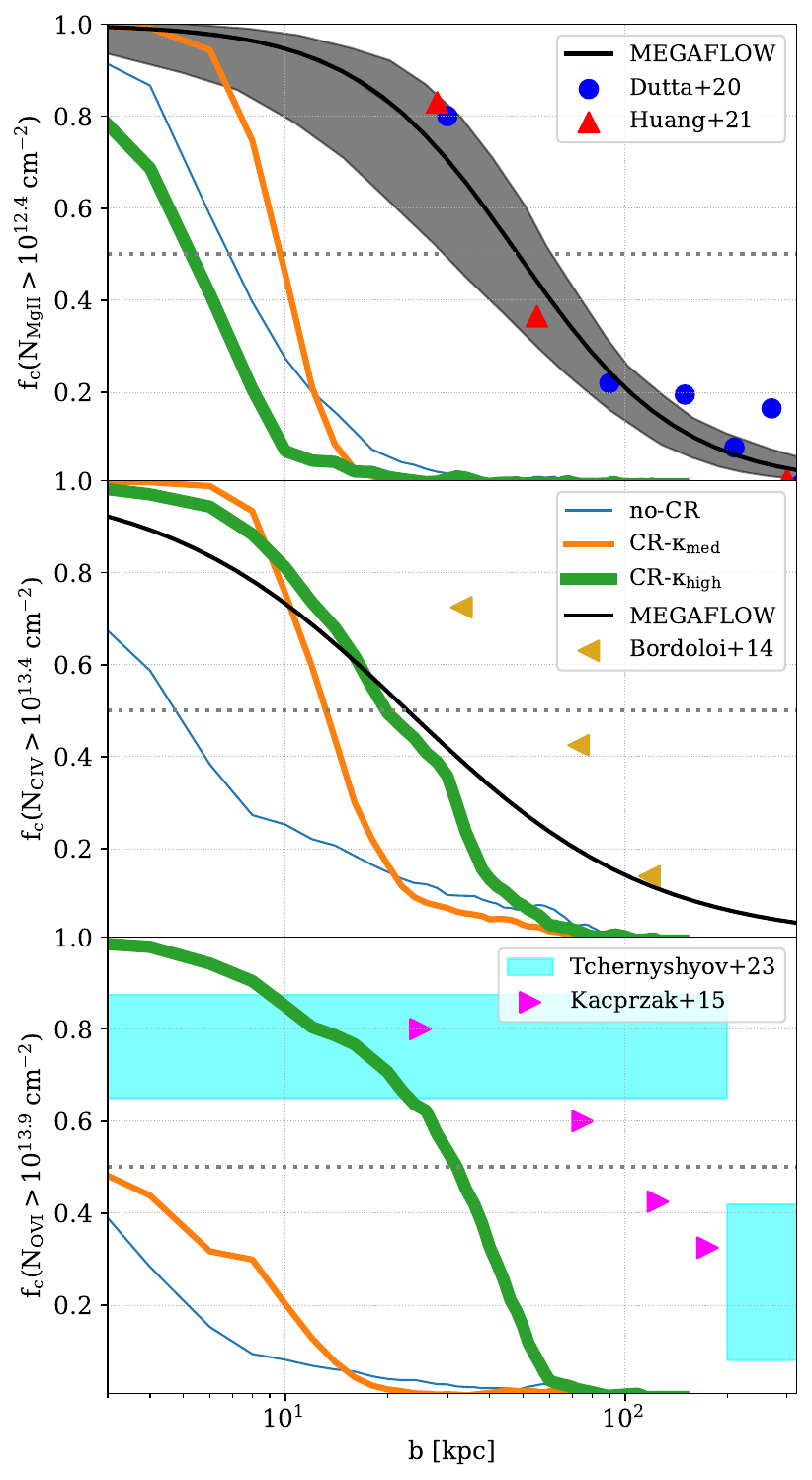} \vspace{-15pt}
\caption{\MgII{} (top), \CIV{} (middle), and \OVI{} (bottom) covering fractions as a function of impact parameter for the three runs, stacked for snapshots with $1 < z < 1.3$. The solid black line and shaded region shows fits and the 95\% confidence region from MEGAFLOW \protect\citep{Schroetter21} observations of \MgII{} and \CIV{} (the 95\% confidence region for \CIV{} is comparable to that of \MgII{}), while the coloured markers and regions show \MgII{} observations from \protect\cite{Dutta20} and \protect\cite{Huang21}, \CIV{} observations from \protect\cite{Bordoloi14}, and \OVI{} observations from \protect\cite{Kacprzak15} and \protect\cite{Tchernyshyov23}. The horizontal gray dotted line in all panels shows a covering fraction of 50\%. All runs, with or without CRs, fail to produce enough \MgII{} or \OVI{} in their CGM to match observations. However, \crhigh{} is effective at boosting the \CIV{} closer to observed levels in the CGM.}
\label{fig:coveringfractions}
\end{figure}

Now, we make an explicit comparison to observations by plotting the covering fraction of the different ions in the CGM of our runs. In Figure \ref{fig:coveringfractions}, we show \MgII, \CIV, and \OVI{} covering fractions as a function of impact parameter for each run calculated using all 12-14 snapshots between $z=1$ and $z=1.3$. This large sample size serves two purposes: first, to increase the number of sightlines used in the calculation and reduce the effect of transient features in the CGM (as in previous figures), and second, to better represent the spread in absorber redshifts in $z\sim1$ surveys like MEGAFLOW. For each snapshot, we choose a random orientation of the halo and measure the column density of sightlines along that orientation with impact parameters as large as the virial radius. We define a sightline to be ``covered'' if it exceeds the ion column density corresponding to an equivalent width threshold used by \cite{Schroetter21} for \MgII{} and \CIV{} and \cite{Kacprzak15} for \OVI. This conversion from equivalent width to column density depends on the wavelength considered and assumes an optically thin regime (see Rey et al. 2023, in prep. for details) and results in minimum column densities of $10^{12.4} \; \rm{cm^{-2}}$ for \MgII, $10^{13.4} \; \rm{cm^{-2}}$ for \CIV, and $10^{13.9} \; \rm{cm^{-2}}$ for \OVI, all roughly corresponding to equivalent widths of 0.1 \AA.

From the upper panel of Figure \ref{fig:coveringfractions}, it is clear that none of the runs produce nearly enough \MgII{} absorption in the CGM to be consistent with observations from MEGAFLOW, as well as with similar observations from other recent \MgII{} surveys from \cite{Dutta20} at $0.8 < z < 1.5$ and \cite{Huang21} at $0.1 < z < 0.5$. Within the galaxy ($< 10$ kpc), the \crmed{} run produces the highest \MgII{} covering fractions, boosting a bit the typical values seen in the \nocr{} run. The \crhigh{} run lowers the covering fraction at these impact parameters. All of the runs drop below a covering fraction of 50\% by 10 kpc rather than at $\approx 50 \; \rm{kpc}$ as in the observations. Outside of the galaxy, it is actually the \nocr{} run that has the highest overall covering fractions, largely coming from the high-column density tidal tails connected to the galaxy that are strongest in that run. However, all of the runs are very \MgII-deficient at these impact parameters.

The middle panel shows covering fractions for \CIV{} as well as comparable observations from \cite{Bordoloi14} and MEGAFLOW \citep{Schroetter21}. All three runs are better at matching observed \CIV{} covering fractions from MEGAFLOW than they are at matching \MgII{} as a function of impact parameter. Within the galaxy, both runs with CRs show an enhancement of the covering fraction. Both \nocr{} and \crmed{} drop below 50\% at impact parameters $< 15$ kpc, noticeably closer to the galaxy than both MEGAFLOW and \cite{Bordoloi14}, and in the CGM both of these runs are significantly below the observed covering fractions. The \crhigh{} run is different: it stays much closer to the observed values from MEGAFLOW until $\approx40$ kpc where it starts to fall short. However, it is still significantly below the lower-redshift observations from \cite{Bordoloi14}.

Finally, we show the three runs' covering fractions for \OVI, as well as comparisons to recent observations from \cite{Kacprzak15} and \cite{Tchernyshyov23}. As for \MgII, the \nocr{} and \crmed{} runs fail to reproduce observable \OVI{} in the CGM. The \crhigh{} run is significantly closer to observations, though at nearly all impact parameters in the CGM that run still falls very short. Interestingly, within the galaxy, only the \crhigh{} run has enough \OVI{} to approach observed values of the covering fraction, likely indicative of the higher metallicity environment of the \crhigh{} run seen in Figure \ref{fig:galaxyoverview}. As with \cite{Bordoloi14}, these two \OVI{} surveys are at lower redshifts than our fiducial simulation outputs. Running the simulations to a matching lower redshift could allow the CRs more time to diffuse out from the galaxy and affect the CGM, resulting in a better agreement between the covering fractions of \CIV{} or \OVI{}. However, this is unlikely to occur as from redshifts $z=1.3$ to $z=1$ none of our simulated covering fractions consistently increase or decrease.

\section{Discussion}
\label{sec:discussion}

In this section, we first consider the effect of varying the column density thresholds used in deriving covering fractions from our simulations, to determine how sensitive our observational comparison is to small (and large) adjustments to these values. Then, we discuss our results in the context of other recent studies on the effects CRs have on the CGM. 

\subsection{Column density cutoffs}

\begin{figure*}
\includegraphics[width=\textwidth]{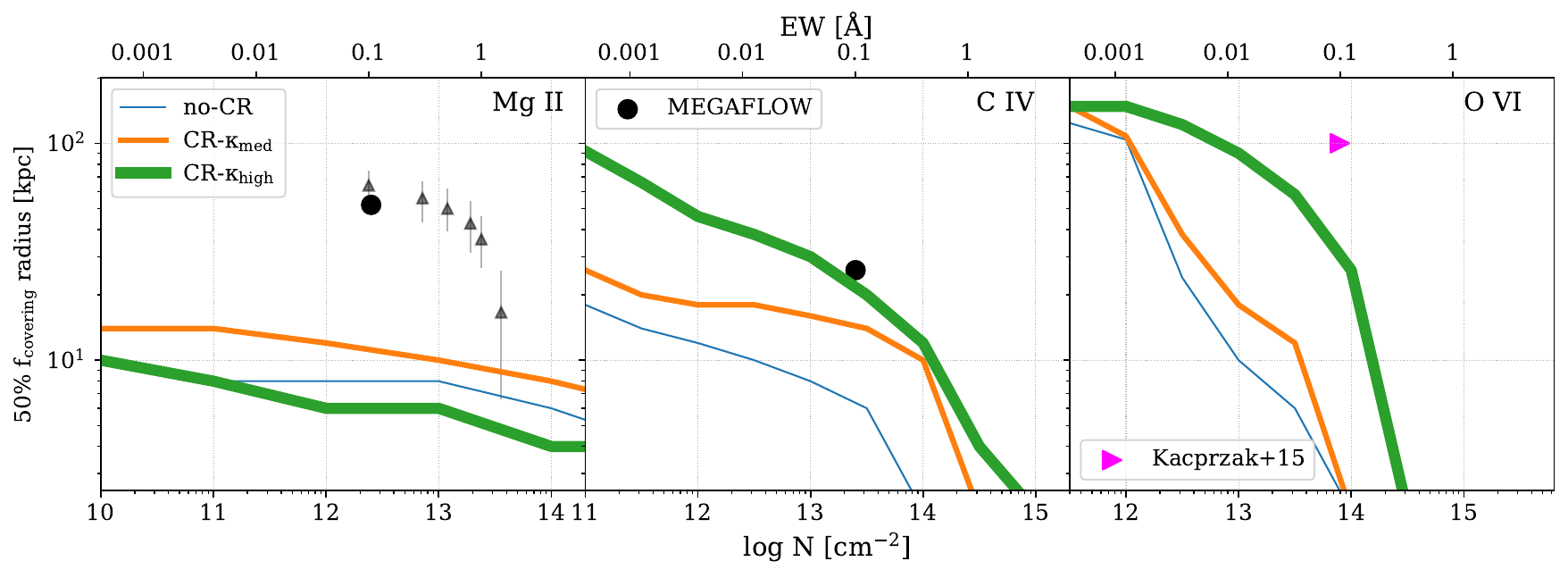} \vspace{-15pt}
\caption{Radius at which \MgII{} (left), \CIV{} (middle), and \OVI{} (right) reach a 50\% covering fraction depending on the column density threshold chosen. Conversions between equivalent width and column density are shown on the top and bottom axes. The two solid black markers represent the observations from MEGAFLOW \citep{Schroetter21} also shown in Figure \ref{fig:coveringfractions}, while the grey points with error bars in the left panel are from the most recent MEGAFLOW analysis by Cherrey et al. in prep.. The pink marker represents the covering radius for \OVI{} from \protect\cite{Kacprzak15}.}
\label{fig:Ncutoff}
\end{figure*}

The main results of our comparison to observations in the previous section come from assuming particular column density thresholds for the different ions derived from recent observational studies and applying those to our simulations. These precise thresholds depend on specific properties of the surveys such as the length of observations, as well as the sensitivity of the actual instruments. We may therefore reach different conclusions if a deeper set of observations of any of these ions are used. To measure this possible effect, we vary the column density threshold used to derive covering fractions, thus mimicking the effect of observing the same object with different sensitivities or resolutions. We quantify this in Figure \ref{fig:Ncutoff}, by plotting the radius at which 50\% of sightlines are higher than a series of thresholds for \MgII{}, \CIV{}, and \OVI{} as a function of that column density threshold. We see that for \MgII{}, this radius shows little evolution over 4 dex in column density for all three runs, and reaches a maximum of only $\approx15$ kpc for a minimum column density of $10^{10} \; \rm{cm^{-2}}$. This is substantially smaller than the corresponding radius of 52 kpc from the fit to the data from \cite{Schroetter21}, as well as the radii calculated by Cherrey et al. in prep. using multiple \MgII{} equivalent width cutoffs $\leq 1$ \AA, indicating that the simulated covering fractions are consistently below observations and not sensitive at all to the precise choice of column density (or equivalent width) used to define observable \MgII{} in our simulations. 

\CIV{} column densities span fewer orders of magnitude throughout the halo, so its 50\% covering fraction radius shows a stronger trend with the column density threshold. Reducing the column density (or equivalent width) cutoff over a common observational range from $\approx 1$ \AA \; to $\approx 0.1$ \AA \; significantly increases the 50\% covering fraction radius by at least half a dex. The \crhigh{} run in particular matches the the 50\% covering fraction radius of MEGAFLOW very near the survey's minimum \CIV{} column density of $\sim10^{13.4} \; \rm{cm^{-2}}$, and these simulations suggest that more sensitive \CIV{} observations would reveal higher covering fractions for that ion that extend to larger and larger radii. Finally, we see that like \CIV{}, the extent of \OVI{} coverage is also very sensitive to the column density threshold. All three of our runs' 50\% covering fraction radii vary in a similar way with column density, and slightly more strongly than in the corresponding \CIV{} panel. The \crhigh{} run in particular has a consistently larger 50\% covering fraction radius at all considered column densities.

The enhancement of \CIV{} and \OVI{} in the \crhigh{} run at multiple column density thresholds compared to the other two runs shows that CRs can significantly contribute to the metal enrichment of the CGM and provide an environment favorable to certain ions, but only if they are able to escape the ISM so their pressure support becomes substantial at large distances. The reason \MgII{} behaves differently in Figure \ref{fig:Ncutoff} is due to the lack of cold ($\leq 10^4 \; \rm{K}$) gas in the CGM rather than a lack of metals, which itself comes from the loss of resolution of the coldest gas structures from the ISM to the CGM, resulting in very low \MgII{} column densities in the CGM. Simulations with increased CGM resolution such as \cite{Hummels19} and \cite{vandeVoort19} suggest that the typical temperature of the CGM should be lower, and heavily favor the formation of low ions like \MgII. Thus, with significantly increased resolution in the CGM, we speculate that effective CR diffusion out of the ISM would produce a similar enhancement in the \MgII{} covering fraction as it currently does for \CIV{} and \OVI{}.

\subsection{Comparisons to recent work}

There have been many recent efforts in adding CRs to galaxy formation simulations, and we highlight some with direct connections to our analysis of galaxy and especially CGM gas properties here. \cite{Farcy22} modelled a set of idealised galaxies using a similar CR feedback implementation, and we compare our results to the most massive galaxy from that study as it is most similar to our simulations. In the ISM, they also find that the gas density distribution is smoother and the SFR is reduced when CRs are included. However, the effect of different diffusion coefficients changes when the galaxies are modelled from cosmological initial conditions. In \cite{Farcy22}'s idealised galaxies, the largest reduction in the SFR occurs for lower stellar masses with $\kappa=10^{27} \; \rm{cm^2 \; s^{-1}}$, and as $\kappa$ increases CRs are less effective at reducing the SFR. At higher stellar masses the SFR is essentially unaffected by CRs regardless of the choice of $\kappa$. In our simulations the trend with $\kappa$ is different: we find that the higher diffusion in \crhigh{} results in a lower SFR over time than \crmed{}. This is likely because our simulations have cosmological inflows which are slowed by CRs that diffuse out of the galaxy more effectively, thus reducing fuel for star formation. We also see a qualitatively similar result in the CGM: gas surrounding the idealised galaxies is cooler when CRs are included. This is driven entirely by the changing nature of the outflowing gas, which is dominated by both warm ($10^4 \; K < T < 10^5 \; K$) and hot ($T > 10^5 \; K$) gas with CRs, and only hot gas without CRs, when measured 10 kpc above the galaxy disc plane. However, in our cosmological simulations, we see that unless the CR diffusivity is high, gas at temperatures below $10^5 \; K$ do not extend very far into the CGM at all (see Figure \ref{fig:CGMoverview}). Furthermore, from Figure \ref{fig:outflow_rates} we see that the highest outflow rates very close to the galaxy are in all cases dominated by cold ($T < 10^{4.5} \; K$) gas, but further out beyond 10 kpc only hot gas is outflowing. Unlike in \cite{Farcy22}, $\kappa=10^{28} \; \rm{cm^2\; s^{-1}}$ as used in \crmed{} results in very little outflow enhancement at any temperature, meaning that in a cosmological environment (i.e. with longer physical timescales and large-scale inflows), a higher level of CR diffusivity is necessary to enhance the outflow rate to distances beyond 10 kpc.

\cite{Rodriguez_Montero23} also study the effect of CR feedback on properties of the ISM and outflows using a \ramses{} simulation setup much more similar to ours: namely, a cosmological zoom-in simulation of a Milky Way-analogue evolved to $z=1.5$. They use a CR diffusion coefficient of $3\times10^{28} \; \rm{cm^2 \; s^{-1}}$, which lies between our \crmed{} and \crhigh{}, although they also include CR streaming. Though we focus our attention on circumgalactic gas, we still find many key consistencies in the effects CRs have on our simulated galaxies. For example, their CR simulation has an early reduction in the stellar mass that levels off to a factor of a few by the end of their simulation, much like what we see in Figure \ref{fig:SFR_stellarmass}, although their inclusion of CR streaming has an impact on their star formation history that is not modelled in our analysis. CRs also smooth out the gas distribution in the disc. Furthermore, they find that CR-launched outflows are more dominated by ``warm'' ($10^4 \; K < T < 10^5 \; K$) gas than outflows without CRs. Particularly relevant to our work, they find that a non-thermal pressure gradient similar to what we find in Figure \ref{fig:pressure} further accelerates outflowing gas in the CGM of their simulation, demonstrating how CRs can redistribute gas on galactic and circumgalactic scales concurrently.

There have also been many direct comparisons to CGM observations using various simulation codes such as Enzo, Gizmo and ChaNGa, which have found differing specific effects. While the overall temperatures of the CGM are cooler with CRs, the column density profiles of key ions do not always change in the same way. For example, in simulations from \cite{Salem16}, \OVI{} column densities are enhanced by nearly a factor of 100 when CRs are included in the physics model. Other studies have less drastic changes: in the inner regions of the CGM, \cite{Ji20} finds an enhancement of the \OVI{} column density by a factor of $\approx3$ for halo masses comparable to ours. However, the simulations from \cite{Butsky22} actually have lower \OVI{} column densities in the CGM with CRs than without. We speculate that this opposite effect observed in \cite{Butsky22} originates from their ``blastwave'' supernova feedback model, in which cooling is temporarily disabled over some timescale. \cite{Rosdahl17} showed that turning off cooling in supernova remnants tends to produce much cooler (i.e. \OVI-poorer) outflows than other supernova feedback models, so adding CRs may not boost \OVI{} column densities in the CGM.

More intermediate ions like \SiIII{} and \SiIV{} are consistently enhanced in these simulations at levels currently probed by observations, but \MgII{} as measured by \cite{Ji20} is only enhanced at column densities that are below current observational limits for radii outside the inner $\approx30$ kpc. Importantly, this indicates that high-equivalent width \MgII{} observations such as those from MEGAFLOW are difficult to reproduce with CRs across multiple simulation codes. The simulations in these two studies use CR diffusion coefficients of 1 to a few $\times \; 10^{29} \; \rm{cm^2\;s^{-1}}$, suggesting that higher values of diffusion are favored for matching CGM observations, especially at larger radii. The comparably high CR diffusion we use in our simulations is not enough to increase \MgII{} to the levels seen in large CGM absorption surveys, although as CR transport is fairly unconstrained the possibility of CR diffusion alone affecting the CGM at those levels cannot be ruled out. It is also likely that with higher resolution of the cold phase, CRs effect on \MgII{} in the CGM could be larger and bring the simulations closer to observations, even with the same CR diffusion coefficient.

\section{Conclusions}
\label{sec:conclusions}

In this paper, we ran three \ramses{} simulations in order to understand the possible range of effects that CR feedback has on the CGM of galaxies. We evolved three realizations of the same galaxy from cosmological initial conditions to $z=1$ with no additional CR feedback (\nocr), CR feedback with a moderate value of $10^{28} \; \rm{cm^2\;s^{-1}}$ for the CR diffusion coefficient (\crmed), and CR feedback with a high value of $3\times10^{29} \; \rm{cm^2\;s^{-1}}$ for the CR diffusion coefficient (\crhigh). Our conclusions are as follows:

\begin{enumerate}
\item Over cosmological time, cosmic rays can smooth out the density distribution within the galaxy's ISM and expand the gas disc, though if the cosmic ray diffusion coefficient $\kappa$ is large, this effect is minimal (Figure \ref{fig:galaxyoverview}). 
\item As is the case for galaxies of a similar mass from other idealised and zoom-in simulations, cosmic rays  lower the star formation rate, resulting in a slightly lower stellar mass by $z=1$ (Figure \ref{fig:SFR_stellarmass}).
\item The \crhigh{} run with a higher cosmic ray diffusion coefficient has a CGM that is cooler and, crucially, much richer in metals than either the \nocr{} run or the \crmed{} run, indicating that the ``sweet spot'' of CR diffusivity \citep[see, e.g.][]{Hopkins20} necessary for CRs to affect the phase of the gas and the observability of metal ions beyond the immediate vicinity of the galaxy without completely decoupling the from the gas is at least $ 10^{29} \; \rm{cm^{2} \; s^{-1}}$ (Figures \ref{fig:CGMoverview} and \ref{fig:phasediagrams}).
\item Cosmic rays with a high diffusion coefficient accelerate outflowing gas substantially further out into the CGM by enhancing such gas with temperatures $T \geq 10^5 \; K$ at distances above $\approx20$ kpc from the galaxy (Figure \ref{fig:outflow_rates}). 
\item Cosmic ray pressure dominates but remains confined to the galaxy for smaller $\kappa$, but it dominates or is comparable to thermal pressure in the entire halo for larger $\kappa$, thus allowing more low temperature gas to exist throughout the halo (Figure \ref{fig:pressure}). 
\item Cosmic rays do not significantly increase \MgII{} column densities anywhere in the halo, although they do restructure \MgII{} found near the galaxy-halo interface. The \crhigh{} run noticeably enhances ions found at higher temperatures (\CIV{} and \OVI{}) throughout the halo (Figure \ref{fig:Nmaps}).
\item All three of our runs fail to match observed \MgII{} covering fractions in the CGM from multiple surveys at $z\approx1$. The \crhigh{} run in particular actually lowers the \MgII{} covering fraction at all impact parameters. However, the same run increases the covering fraction of \CIV{} and brings it more in line with MEGAFLOW observations (Figure \ref{fig:coveringfractions}).
\item By reducing the column density threshold used to define the covering fraction, the ``observed'' extent of \CIV{} and \OVI{} moves outwards into the CGM, especially for the \crhigh{} run. This does not occur for \MgII{} because its spread in column densities between small and large radii is much bigger than the same spread for the other measured ions (Figure \ref{fig:Ncutoff}).
\end{enumerate}

With this work, we have studied how CR feedback can propagate out from the galaxy and affect the CGM differently depending on the CR diffusion coefficient. While the \MgII{} content of the CGM appears largely unaffected by the addition of CRs, the CGM as a whole and outflowing gas in particular have a lower temperature when CRs are able to effectively diffuse out from the galaxy. This diffusion is more relevant when modelling outflows with a cosmological zoom-in simulation rather than from an idealised galaxy without any inflows. We expect CR diffusion to be even more effective in future high-resolution studies of the CGM where there will likely be a more prominent cold phase for CRs to influence.

We note that the CRs in this study all propagate with a constant rate of diffusion. Recent work has focused on a more realistic treatment of CR transport by allowing $\kappa$ to vary with gas properties \citep{Farber18,Semenov21}, or by modelling the CR spectrum which allows $\kappa$ to vary with CR energy \citep{Girichidis22}. \cite{Butsky23} confirms that a constant $\kappa$ cannot reproduce the observed complexity of the CGM of COS-Halos galaxies, showing that these more detailed models are indeed necessary for future work. Additionally, other CR transport methods we have not modelled such as streaming could significantly change how the energy from CRs affects the temperature and density structures found in the CGM \citep{Butsky18}, and the evolution of the galaxy in general \citep{Wiener17}. The importance of CR streaming relative to diffusion is an active area of study as well \citep[e.g.][]{Thomas23}.

As work on this topic continues, we intend to further examine the possible constructive impact on CRs of other physical effects (e.g., radiative transfer) and sources of feedback (e.g., AGN) that are not included in our simulations, as this may help provide the physical coupling necessary to produce cold \MgII-bearing outflows that are found in observations.

\section*{Acknowledgements}

We thank the anonymous referee for helpful comments that improved the paper. This work has been carried out thanks to the support of the ANR 3DGasFlows (ANR-17-CE31-0017). Simulations were run using the GENCI allocations A0070410560 from 2019 and A0070410560 from 2020, and they were stored and analysed on PSMN (P\^ole Scientifique de Mod\'elisation Num\'erique) of the ENS de Lyon.

\section*{Data Availability}

The data generated and used in this article will be shared on reasonable request to the corresponding author.



\bibliographystyle{mnras}
\bibliography{references} 

\begin{thebibliography}{}
\makeatletter
\relax
\def\mn@urlcharsother{\let\do\@makeother \do\$\do\&\do\#\do\^\do\_\do\%\do\~}
\def\mn@doi{\begingroup\mn@urlcharsother \@ifnextchar [ {\mn@doi@}
  {\mn@doi@[]}}
\def\mn@doi@[#1]#2{\def\@tempa{#1}\ifx\@tempa\@empty \href
  {http://dx.doi.org/#2} {doi:#2}\else \href {http://dx.doi.org/#2} {#1}\fi
  \endgroup}
\def\mn@eprint#1#2{\mn@eprint@#1:#2::\@nil}
\def\mn@eprint@arXiv#1{\href {http://arxiv.org/abs/#1} {{\tt arXiv:#1}}}
\def\mn@eprint@dblp#1{\href {http://dblp.uni-trier.de/rec/bibtex/#1.xml}
  {dblp:#1}}
\def\mn@eprint@#1:#2:#3:#4\@nil{\def\@tempa {#1}\def\@tempb {#2}\def\@tempc
  {#3}\ifx \@tempc \@empty \let \@tempc \@tempb \let \@tempb \@tempa \fi \ifx
  \@tempb \@empty \def\@tempb {arXiv}\fi \@ifundefined
  {mn@eprint@\@tempb}{\@tempb:\@tempc}{\expandafter \expandafter \csname
  mn@eprint@\@tempb\endcsname \expandafter{\@tempc}}}

\bibitem[\protect\citeauthoryear{{Aubert}, {Pichon}  \& {Colombi}}{{Aubert}
  et~al.}{2004}]{Aubert04}
{Aubert} D.,  {Pichon} C.,   {Colombi} S.,  2004, \mn@doi [\mnras]
  {10.1111/j.1365-2966.2004.07883.x}, \href
  {https://ui.adsabs.harvard.edu/abs/2004MNRAS.352..376A} {352, 376}

\bibitem[\protect\citeauthoryear{{Avery} et~al.,}{{Avery}
  et~al.}{2022}]{Avery22}
{Avery} C.~R.,  et~al., 2022, \mn@doi [\mnras] {10.1093/mnras/stac190}, \href
  {https://ui.adsabs.harvard.edu/abs/2022MNRAS.511.4223A} {511, 4223}

\bibitem[\protect\citeauthoryear{{Bacon} et~al.,}{{Bacon}
  et~al.}{2010}]{Bacon10}
{Bacon} R.,  et~al., 2010, in {McLean} I.~S.,  {Ramsay} S.~K.,   {Takami} H.,
  eds,  Society of Photo-Optical Instrumentation Engineers (SPIE) Conference
  Series Vol. 7735, Ground-based and Airborne Instrumentation for Astronomy
  III. p. 773508 (\mn@eprint {arXiv} {2211.16795}), \mn@doi{10.1117/12.856027}

\bibitem[\protect\citeauthoryear{{Boogaard} et~al.,}{{Boogaard}
  et~al.}{2018}]{Boogaard18}
{Boogaard} L.~A.,  et~al., 2018, \mn@doi [\aap] {10.1051/0004-6361/201833136},
  \href {https://ui.adsabs.harvard.edu/abs/2018A&A...619A..27B} {619, A27}

\bibitem[\protect\citeauthoryear{{Bordoloi} et~al.,}{{Bordoloi}
  et~al.}{2014}]{Bordoloi14}
{Bordoloi} R.,  et~al., 2014, \mn@doi [\apj] {10.1088/0004-637X/796/2/136},
  \href {https://ui.adsabs.harvard.edu/abs/2014ApJ...796..136B} {796, 136}

\bibitem[\protect\citeauthoryear{{Boulares} \& {Cox}}{{Boulares} \&
  {Cox}}{1990}]{Boulares90}
{Boulares} A.,  {Cox} D.~P.,  1990, \mn@doi [\apj] {10.1086/169509}, \href
  {https://ui.adsabs.harvard.edu/abs/1990ApJ...365..544B} {365, 544}

\bibitem[\protect\citeauthoryear{{Burchett} et~al.,}{{Burchett}
  et~al.}{2019}]{Burchett19}
{Burchett} J.~N.,  et~al., 2019, \mn@doi [\apjl] {10.3847/2041-8213/ab1f7f},
  \href {https://ui.adsabs.harvard.edu/abs/2019ApJ...877L..20B} {877, L20}

\bibitem[\protect\citeauthoryear{{Butsky} \& {Quinn}}{{Butsky} \&
  {Quinn}}{2018}]{Butsky18}
{Butsky} I.~S.,  {Quinn} T.~R.,  2018, \mn@doi [\apj]
  {10.3847/1538-4357/aaeac2}, \href
  {https://ui.adsabs.harvard.edu/abs/2018ApJ...868..108B} {868, 108}

\bibitem[\protect\citeauthoryear{{Butsky} et~al.,}{{Butsky}
  et~al.}{2022}]{Butsky22}
{Butsky} I.~S.,  et~al., 2022, \mn@doi [\apj] {10.3847/1538-4357/ac7ebd}, \href
  {https://ui.adsabs.harvard.edu/abs/2022ApJ...935...69B} {935, 69}

\bibitem[\protect\citeauthoryear{{Butsky}, {Nakum}, {Ponnada}, {Hummels}, {Ji}
  \& {Hopkins}}{{Butsky} et~al.}{2023}]{Butsky23}
{Butsky} I.~S.,  {Nakum} S.,  {Ponnada} S.~B.,  {Hummels} C.~B.,  {Ji} S.,
  {Hopkins} P.~F.,  2023, \mn@doi [\mnras] {10.1093/mnras/stad671}, \href
  {https://ui.adsabs.harvard.edu/abs/2023MNRAS.521.2477B} {521, 2477}

\bibitem[\protect\citeauthoryear{{Chan}, {Kere{\v{s}}}, {Hopkins}, {Quataert},
  {Su}, {Hayward}  \& {Faucher-Gigu{\`e}re}}{{Chan} et~al.}{2019}]{Chan19}
{Chan} T.~K.,  {Kere{\v{s}}} D.,  {Hopkins} P.~F.,  {Quataert} E.,  {Su} K.~Y.,
   {Hayward} C.~C.,   {Faucher-Gigu{\`e}re} C.~A.,  2019, \mn@doi [\mnras]
  {10.1093/mnras/stz1895}, \href
  {https://ui.adsabs.harvard.edu/abs/2019MNRAS.488.3716C} {488, 3716}

\bibitem[\protect\citeauthoryear{{Chisholm}, {Bordoloi}, {Rigby}  \&
  {Bayliss}}{{Chisholm} et~al.}{2018}]{Chisholm18}
{Chisholm} J.,  {Bordoloi} R.,  {Rigby} J.~R.,   {Bayliss} M.,  2018, \mn@doi
  [\mnras] {10.1093/mnras/stx2848}, \href
  {https://ui.adsabs.harvard.edu/abs/2018MNRAS.474.1688C} {474, 1688}

\bibitem[\protect\citeauthoryear{{Dashyan} \& {Dubois}}{{Dashyan} \&
  {Dubois}}{2020}]{Dashyan20}
{Dashyan} G.,  {Dubois} Y.,  2020, \mn@doi [\aap]
  {10.1051/0004-6361/201936339}, \href
  {https://ui.adsabs.harvard.edu/abs/2020A&A...638A.123D} {638, A123}

\bibitem[\protect\citeauthoryear{{Davies}, {Crain}, {Oppenheimer}  \&
  {Schaye}}{{Davies} et~al.}{2020}]{Davies20}
{Davies} J.~J.,  {Crain} R.~A.,  {Oppenheimer} B.~D.,   {Schaye} J.,  2020,
  \mn@doi [\mnras] {10.1093/mnras/stz3201}, \href
  {https://ui.adsabs.harvard.edu/abs/2020MNRAS.491.4462D} {491, 4462}

\bibitem[\protect\citeauthoryear{{DeFelippis}, {Genel}, {Bryan}  \&
  {Fall}}{{DeFelippis} et~al.}{2017}]{DeFelippis17}
{DeFelippis} D.,  {Genel} S.,  {Bryan} G.~L.,   {Fall} S.~M.,  2017, \mn@doi
  [\apj] {10.3847/1538-4357/aa6dfc}, \href
  {https://ui.adsabs.harvard.edu/abs/2017ApJ...841...16D} {841, 16}

\bibitem[\protect\citeauthoryear{{Dubois} \& {Commer{\c{c}}on}}{{Dubois} \&
  {Commer{\c{c}}on}}{2016}]{Dubois16}
{Dubois} Y.,  {Commer{\c{c}}on} B.,  2016, \mn@doi [\aap]
  {10.1051/0004-6361/201527126}, \href
  {https://ui.adsabs.harvard.edu/abs/2016A&A...585A.138D} {585, A138}

\bibitem[\protect\citeauthoryear{{Dubois}, {Peirani}, {Pichon}, {Devriendt},
  {Gavazzi}, {Welker}  \& {Volonteri}}{{Dubois} et~al.}{2016}]{Dubois16a}
{Dubois} Y.,  {Peirani} S.,  {Pichon} C.,  {Devriendt} J.,  {Gavazzi} R.,
  {Welker} C.,   {Volonteri} M.,  2016, \mn@doi [\mnras]
  {10.1093/mnras/stw2265}, \href
  {https://ui.adsabs.harvard.edu/abs/2016MNRAS.463.3948D} {463, 3948}

\bibitem[\protect\citeauthoryear{{Dubois}, {Commer{\c{c}}on}, {Marcowith}  \&
  {Brahimi}}{{Dubois} et~al.}{2019}]{Dubois19}
{Dubois} Y.,  {Commer{\c{c}}on} B.,  {Marcowith} A.,   {Brahimi} L.,  2019,
  \mn@doi [\aap] {10.1051/0004-6361/201936275}, \href
  {https://ui.adsabs.harvard.edu/abs/2019A&A...631A.121D} {631, A121}

\bibitem[\protect\citeauthoryear{{Dutta} et~al.,}{{Dutta}
  et~al.}{2020}]{Dutta20}
{Dutta} R.,  et~al., 2020, \mn@doi [\mnras] {10.1093/mnras/staa3147}, \href
  {https://ui.adsabs.harvard.edu/abs/2020MNRAS.499.5022D} {499, 5022}

\bibitem[\protect\citeauthoryear{{Farber}, {Ruszkowski}, {Yang}  \&
  {Zweibel}}{{Farber} et~al.}{2018}]{Farber18}
{Farber} R.,  {Ruszkowski} M.,  {Yang} H. Y.~K.,   {Zweibel} E.~G.,  2018,
  \mn@doi [\apj] {10.3847/1538-4357/aab26d}, \href
  {https://ui.adsabs.harvard.edu/abs/2018ApJ...856..112F} {856, 112}

\bibitem[\protect\citeauthoryear{{Farcy}, {Rosdahl}, {Dubois}, {Blaizot}  \&
  {Martin-Alvarez}}{{Farcy} et~al.}{2022}]{Farcy22}
{Farcy} M.,  {Rosdahl} J.,  {Dubois} Y.,  {Blaizot} J.,   {Martin-Alvarez} S.,
  2022, \mn@doi [\mnras] {10.1093/mnras/stac1196}, \href
  {https://ui.adsabs.harvard.edu/abs/2022MNRAS.513.5000F} {513, 5000}

\bibitem[\protect\citeauthoryear{{Faucher-Gigu{\`e}re} \&
  {Oh}}{{Faucher-Gigu{\`e}re} \& {Oh}}{2023}]{Faucher-Giguere23}
{Faucher-Gigu{\`e}re} C.-A.,  {Oh} S.~P.,  2023, \mn@doi [\araa]
  {10.1146/annurev-astro-052920-125203}, \href
  {https://ui.adsabs.harvard.edu/abs/2023ARA&A..61..131F} {61, 131}

\bibitem[\protect\citeauthoryear{{Ferland}, {Korista}, {Verner}, {Ferguson},
  {Kingdon}  \& {Verner}}{{Ferland} et~al.}{1998}]{Ferland98}
{Ferland} G.~J.,  {Korista} K.~T.,  {Verner} D.~A.,  {Ferguson} J.~W.,
  {Kingdon} J.~B.,   {Verner} E.~M.,  1998, \mn@doi [\pasp] {10.1086/316190},
  \href {https://ui.adsabs.harvard.edu/abs/1998PASP..110..761F} {110, 761}

\bibitem[\protect\citeauthoryear{{Ferland} et~al.,}{{Ferland}
  et~al.}{2013}]{Ferland13}
{Ferland} G.~J.,  et~al., 2013, \rmxaa, \href
  {https://ui.adsabs.harvard.edu/abs/2013RMxAA..49..137F} {49, 137}

\bibitem[\protect\citeauthoryear{{Girichidis}, {Naab}, {Hanasz}  \&
  {Walch}}{{Girichidis} et~al.}{2018}]{Girichidis18}
{Girichidis} P.,  {Naab} T.,  {Hanasz} M.,   {Walch} S.,  2018, \mn@doi
  [\mnras] {10.1093/mnras/sty1653}, \href
  {https://ui.adsabs.harvard.edu/abs/2018MNRAS.479.3042G} {479, 3042}

\bibitem[\protect\citeauthoryear{{Girichidis}, {Pfrommer}, {Pakmor}  \&
  {Springel}}{{Girichidis} et~al.}{2022}]{Girichidis22}
{Girichidis} P.,  {Pfrommer} C.,  {Pakmor} R.,   {Springel} V.,  2022, \mn@doi
  [\mnras] {10.1093/mnras/stab3462}, \href
  {https://ui.adsabs.harvard.edu/abs/2022MNRAS.510.3917G} {510, 3917}

\bibitem[\protect\citeauthoryear{{Girichidis}, {Werhahn}, {Pfrommer}, {Pakmor}
  \& {Springel}}{{Girichidis} et~al.}{2024}]{Girichidis24}
{Girichidis} P.,  {Werhahn} M.,  {Pfrommer} C.,  {Pakmor} R.,   {Springel} V.,
  2024, \mn@doi [\mnras] {10.1093/mnras/stad3628}, \href
  {https://ui.adsabs.harvard.edu/abs/2024MNRAS.527.10897} {527, 10897}

\bibitem[\protect\citeauthoryear{{Guo} \& {Oh}}{{Guo} \& {Oh}}{2008}]{Gu08}
{Guo} F.,  {Oh} S.~P.,  2008, \mn@doi [\mnras]
  {10.1111/j.1365-2966.2007.12692.x}, \href
  {https://ui.adsabs.harvard.edu/abs/2008MNRAS.384..251G} {384, 251}

\bibitem[\protect\citeauthoryear{{Haardt} \& {Madau}}{{Haardt} \&
  {Madau}}{1996}]{Haardt96}
{Haardt} F.,  {Madau} P.,  1996, \mn@doi [\apj] {10.1086/177035}, \href
  {https://ui.adsabs.harvard.edu/abs/1996ApJ...461...20H} {461, 20}

\bibitem[\protect\citeauthoryear{{Hahn} \& {Abel}}{{Hahn} \&
  {Abel}}{2011}]{Hahn11}
{Hahn} O.,  {Abel} T.,  2011, \mn@doi [\mnras]
  {10.1111/j.1365-2966.2011.18820.x}, \href
  {https://ui.adsabs.harvard.edu/abs/2011MNRAS.415.2101H} {415, 2101}

\bibitem[\protect\citeauthoryear{{Hopkins} et~al.,}{{Hopkins}
  et~al.}{2020}]{Hopkins20}
{Hopkins} P.~F.,  et~al., 2020, \mn@doi [\mnras] {10.1093/mnras/stz3321}, \href
  {https://ui.adsabs.harvard.edu/abs/2020MNRAS.492.3465H} {492, 3465}

\bibitem[\protect\citeauthoryear{{Hopkins}, {Chan}, {Squire}, {Quataert}, {Ji},
  {Kere{\v{s}}}  \& {Faucher-Gigu{\`e}re}}{{Hopkins} et~al.}{2021}]{Hopkins21}
{Hopkins} P.~F.,  {Chan} T.~K.,  {Squire} J.,  {Quataert} E.,  {Ji} S.,
  {Kere{\v{s}}} D.,   {Faucher-Gigu{\`e}re} C.-A.,  2021, \mn@doi [\mnras]
  {10.1093/mnras/staa3692}, \href
  {https://ui.adsabs.harvard.edu/abs/2021MNRAS.501.3663H} {501, 3663}

\bibitem[\protect\citeauthoryear{{Hopkins}, {Squire}  \& {Butsky}}{{Hopkins}
  et~al.}{2022}]{Hopkins22}
{Hopkins} P.~F.,  {Squire} J.,   {Butsky} I.~S.,  2022, \mn@doi [\mnras]
  {10.1093/mnras/stab2635}, \href
  {https://ui.adsabs.harvard.edu/abs/2022MNRAS.509.3779H} {509, 3779}

\bibitem[\protect\citeauthoryear{{Huang}, {Chen}, {Shectman}, {Johnson},
  {Zahedy}, {Helsby}, {Gauthier}  \& {Thompson}}{{Huang}
  et~al.}{2021}]{Huang21}
{Huang} Y.-H.,  {Chen} H.-W.,  {Shectman} S.~A.,  {Johnson} S.~D.,  {Zahedy}
  F.~S.,  {Helsby} J.~E.,  {Gauthier} J.-R.,   {Thompson} I.~B.,  2021, \mn@doi
  [\mnras] {10.1093/mnras/stab360}, \href
  {https://ui.adsabs.harvard.edu/abs/2021MNRAS.502.4743H} {502, 4743}

\bibitem[\protect\citeauthoryear{{Hummels}, {Smith}  \& {Silvia}}{{Hummels}
  et~al.}{2017}]{Hummels17}
{Hummels} C.~B.,  {Smith} B.~D.,   {Silvia} D.~W.,  2017, \mn@doi [\apj]
  {10.3847/1538-4357/aa7e2d}, \href
  {https://ui.adsabs.harvard.edu/abs/2017ApJ...847...59H} {847, 59}

\bibitem[\protect\citeauthoryear{{Hummels} et~al.,}{{Hummels}
  et~al.}{2019}]{Hummels19}
{Hummels} C.~B.,  et~al., 2019, \mn@doi [\apj] {10.3847/1538-4357/ab378f},
  \href {https://ui.adsabs.harvard.edu/abs/2019ApJ...882..156H} {882, 156}

\bibitem[\protect\citeauthoryear{{Jacob}, {Pakmor}, {Simpson}, {Springel}  \&
  {Pfrommer}}{{Jacob} et~al.}{2018}]{Jacob18}
{Jacob} S.,  {Pakmor} R.,  {Simpson} C.~M.,  {Springel} V.,   {Pfrommer} C.,
  2018, \mn@doi [\mnras] {10.1093/mnras/stx3221}, \href
  {https://ui.adsabs.harvard.edu/abs/2018MNRAS.475..570J} {475, 570}

\bibitem[\protect\citeauthoryear{{Ji} et~al.,}{{Ji} et~al.}{2020}]{Ji20}
{Ji} S.,  et~al., 2020, \mn@doi [\mnras] {10.1093/mnras/staa1849}, \href
  {https://ui.adsabs.harvard.edu/abs/2020MNRAS.496.4221J} {496, 4221}

\bibitem[\protect\citeauthoryear{{Jiang} \& {Oh}}{{Jiang} \&
  {Oh}}{2018}]{Jiang18}
{Jiang} Y.-F.,  {Oh} S.~P.,  2018, \mn@doi [\apj] {10.3847/1538-4357/aaa6ce},
  \href {https://ui.adsabs.harvard.edu/abs/2018ApJ...854....5J} {854, 5}

\bibitem[\protect\citeauthoryear{{Kacprzak}, {Muzahid}, {Churchill}, {Nielsen}
  \& {Charlton}}{{Kacprzak} et~al.}{2015}]{Kacprzak15}
{Kacprzak} G.~G.,  {Muzahid} S.,  {Churchill} C.~W.,  {Nielsen} N.~M.,
  {Charlton} J.~C.,  2015, \mn@doi [\apj] {10.1088/0004-637X/815/1/22}, \href
  {https://ui.adsabs.harvard.edu/abs/2015ApJ...815...22K} {815, 22}

\bibitem[\protect\citeauthoryear{{Kimm} \& {Cen}}{{Kimm} \&
  {Cen}}{2014}]{Kimm14}
{Kimm} T.,  {Cen} R.,  2014, \mn@doi [\apj] {10.1088/0004-637X/788/2/121},
  \href {https://ui.adsabs.harvard.edu/abs/2014ApJ...788..121K} {788, 121}

\bibitem[\protect\citeauthoryear{{Kimm}, {Cen}, {Devriendt}, {Dubois}  \&
  {Slyz}}{{Kimm} et~al.}{2015}]{Kimm15}
{Kimm} T.,  {Cen} R.,  {Devriendt} J.,  {Dubois} Y.,   {Slyz} A.,  2015,
  \mn@doi [\mnras] {10.1093/mnras/stv1211}, \href
  {https://ui.adsabs.harvard.edu/abs/2015MNRAS.451.2900K} {451, 2900}

\bibitem[\protect\citeauthoryear{{Kopenhafer}, {O'Shea}  \&
  {Voit}}{{Kopenhafer} et~al.}{2023}]{Kopenhafer23}
{Kopenhafer} C.,  {O'Shea} B.~W.,   {Voit} G.~M.,  2023, \mn@doi [\apj]
  {10.3847/1538-4357/accbb7}, \href
  {https://ui.adsabs.harvard.edu/abs/2023ApJ...951..107K} {951, 107}

\bibitem[\protect\citeauthoryear{{Kroupa}}{{Kroupa}}{2001}]{Kroupa01}
{Kroupa} P.,  2001, \mn@doi [\mnras] {10.1046/j.1365-8711.2001.04022.x}, \href
  {https://ui.adsabs.harvard.edu/abs/2001MNRAS.322..231K} {322, 231}

\bibitem[\protect\citeauthoryear{{Marinacci} et~al.,}{{Marinacci}
  et~al.}{2018}]{Marinacci18}
{Marinacci} F.,  et~al., 2018, \mn@doi [\mnras] {10.1093/mnras/sty2206}, \href
  {https://ui.adsabs.harvard.edu/abs/2018MNRAS.480.5113M} {480, 5113}

\bibitem[\protect\citeauthoryear{{Martin-Alvarez}, {Sijacki}, {Haehnelt},
  {Farcy}, {Dubois}, {Belokurov}, {Rosdahl}  \&
  {Lopez-Rodriguez}}{{Martin-Alvarez} et~al.}{2023}]{Martin-Alvarez23}
{Martin-Alvarez} S.,  {Sijacki} D.,  {Haehnelt} M.~G.,  {Farcy} M.,  {Dubois}
  Y.,  {Belokurov} V.,  {Rosdahl} J.,   {Lopez-Rodriguez} E.,  2023, \mn@doi
  [\mnras] {10.1093/mnras/stad2559}, \href
  {https://ui.adsabs.harvard.edu/abs/2023MNRAS.525.3806M} {525, 3806}

\bibitem[\protect\citeauthoryear{{McCourt}, {Oh}, {O'Leary}  \&
  {Madigan}}{{McCourt} et~al.}{2018}]{McCourt18}
{McCourt} M.,  {Oh} S.~P.,  {O'Leary} R.,   {Madigan} A.-M.,  2018, \mn@doi
  [\mnras] {10.1093/mnras/stx2687}, \href
  {https://ui.adsabs.harvard.edu/abs/2018MNRAS.473.5407M} {473, 5407}

\bibitem[\protect\citeauthoryear{{Mitchell}, {Schaye}, {Bower}  \&
  {Crain}}{{Mitchell} et~al.}{2020}]{Mitchell20}
{Mitchell} P.~D.,  {Schaye} J.,  {Bower} R.~G.,   {Crain} R.~A.,  2020, \mn@doi
  [\mnras] {10.1093/mnras/staa938}, \href
  {https://ui.adsabs.harvard.edu/abs/2020MNRAS.494.3971M} {494, 3971}

\bibitem[\protect\citeauthoryear{{Miyoshi} \& {Kusano}}{{Miyoshi} \&
  {Kusano}}{2005}]{Miyoshi05}
{Miyoshi} T.,  {Kusano} K.,  2005, \mn@doi [Journal of Computational Physics]
  {10.1016/j.jcp.2005.02.017}, \href
  {https://ui.adsabs.harvard.edu/abs/2005JCoPh.208..315M} {208, 315}

\bibitem[\protect\citeauthoryear{{Morlino} \& {Caprioli}}{{Morlino} \&
  {Caprioli}}{2012}]{Morlino12}
{Morlino} G.,  {Caprioli} D.,  2012, \mn@doi [\aap]
  {10.1051/0004-6361/201117855}, \href
  {https://ui.adsabs.harvard.edu/abs/2012A&A...538A..81M} {538, A81}

\bibitem[\protect\citeauthoryear{{Naiman} et~al.,}{{Naiman}
  et~al.}{2018}]{Naiman18}
{Naiman} J.~P.,  et~al., 2018, \mn@doi [\mnras] {10.1093/mnras/sty618}, \href
  {https://ui.adsabs.harvard.edu/abs/2018MNRAS.477.1206N} {477, 1206}

\bibitem[\protect\citeauthoryear{{Nelson} et~al.,}{{Nelson}
  et~al.}{2018}]{Nelson18}
{Nelson} D.,  et~al., 2018, \mn@doi [\mnras] {10.1093/mnras/stx3040}, \href
  {https://ui.adsabs.harvard.edu/abs/2018MNRAS.475..624N} {475, 624}

\bibitem[\protect\citeauthoryear{{Nelson} et~al.,}{{Nelson}
  et~al.}{2019}]{Nelson19}
{Nelson} D.,  et~al., 2019, \mn@doi [\mnras] {10.1093/mnras/stz2306}, \href
  {https://ui.adsabs.harvard.edu/abs/2019MNRAS.490.3234N} {490, 3234}

\bibitem[\protect\citeauthoryear{{Nelson} et~al.,}{{Nelson}
  et~al.}{2020}]{Nelson20}
{Nelson} D.,  et~al., 2020, \mn@doi [\mnras] {10.1093/mnras/staa2419}, \href
  {https://ui.adsabs.harvard.edu/abs/2020MNRAS.498.2391N} {498, 2391}

\bibitem[\protect\citeauthoryear{{Nu{\~n}ez-Casti{\~n}eyra}, {Grenier},
  {Bournaud}, {Dubois}, {Kamal Youssef}  \&
  {Hennebelle}}{{Nu{\~n}ez-Casti{\~n}eyra} et~al.}{2022}]{Nunez-Castineyra22}
{Nu{\~n}ez-Casti{\~n}eyra} A.,  {Grenier} I.~A.,  {Bournaud} F.,  {Dubois} Y.,
  {Kamal Youssef} F.~R.,   {Hennebelle} P.,  2022, \mn@doi [arXiv e-prints]
  {10.48550/arXiv.2205.08163}, \href
  {https://ui.adsabs.harvard.edu/abs/2022arXiv220508163N} {p. arXiv:2205.08163}

\bibitem[\protect\citeauthoryear{{Obreja}, {Battaia}, {Macci{\`o}}  \&
  {Buck}}{{Obreja} et~al.}{2023}]{Obreja23}
{Obreja} A.,  {Battaia} F.~A.,  {Macci{\`o}} A.~V.,   {Buck} T.,  2023, \mn@doi
  [\mnras] {10.1093/mnras/stad3410}, \href
  {https://ui.adsabs.harvard.edu/abs/2023MNRAS.tmp.3268O} {}

\bibitem[\protect\citeauthoryear{{Peeples} et~al.,}{{Peeples}
  et~al.}{2019}]{Peeples19}
{Peeples} M.~S.,  et~al., 2019, \mn@doi [\apj] {10.3847/1538-4357/ab0654},
  \href {https://ui.adsabs.harvard.edu/abs/2019ApJ...873..129P} {873, 129}

\bibitem[\protect\citeauthoryear{{Pillepich} et~al.,}{{Pillepich}
  et~al.}{2018a}]{Pillepich18a}
{Pillepich} A.,  et~al., 2018a, \mn@doi [\mnras] {10.1093/mnras/stx2656}, \href
  {https://ui.adsabs.harvard.edu/abs/2018MNRAS.473.4077P} {473, 4077}

\bibitem[\protect\citeauthoryear{{Pillepich} et~al.,}{{Pillepich}
  et~al.}{2018b}]{Pillepich18}
{Pillepich} A.,  et~al., 2018b, \mn@doi [\mnras] {10.1093/mnras/stx3112}, \href
  {https://ui.adsabs.harvard.edu/abs/2018MNRAS.475..648P} {475, 648}

\bibitem[\protect\citeauthoryear{{Pillepich} et~al.,}{{Pillepich}
  et~al.}{2019}]{Pillepich19}
{Pillepich} A.,  et~al., 2019, \mn@doi [\mnras] {10.1093/mnras/stz2338}, \href
  {https://ui.adsabs.harvard.edu/abs/2019MNRAS.490.3196P} {490, 3196}

\bibitem[\protect\citeauthoryear{{Planck Collaboration} et~al.,}{{Planck
  Collaboration} et~al.}{2014}]{Planck14}
{Planck Collaboration} et~al., 2014, \mn@doi [\aap]
  {10.1051/0004-6361/201321591}, \href
  {https://ui.adsabs.harvard.edu/abs/2014A&A...571A..16P} {571, A16}

\bibitem[\protect\citeauthoryear{{Ramesh} \& {Nelson}}{{Ramesh} \&
  {Nelson}}{2024}]{Ramesh24}
{Ramesh} R.,  {Nelson} D.,  2024, \mn@doi [\mnras] {10.1093/mnras/stae237},
  \href {https://ui.adsabs.harvard.edu/abs/2024MNRAS.528.3320R} {528, 3320}

\bibitem[\protect\citeauthoryear{{Rasera} \& {Teyssier}}{{Rasera} \&
  {Teyssier}}{2006}]{Rasera06}
{Rasera} Y.,  {Teyssier} R.,  2006, \mn@doi [\aap]
  {10.1051/0004-6361:20053116}, \href
  {https://ui.adsabs.harvard.edu/abs/2006A&A...445....1R} {445, 1}

\bibitem[\protect\citeauthoryear{{Rey}}{{Rey}}{2022}]{Rey22}
{Rey} M.,  2022, PhD thesis, Ecole Normale Superieure de Lyon, France

\bibitem[\protect\citeauthoryear{{Rodr{\'\i}guez Montero}, {Martin-Alvarez},
  {Slyz}, {Devriendt}, {Dubois}  \& {Sijacki}}{{Rodr{\'\i}guez Montero}
  et~al.}{2023}]{Rodriguez_Montero23}
{Rodr{\'\i}guez Montero} F.,  {Martin-Alvarez} S.,  {Slyz} A.,  {Devriendt} J.,
   {Dubois} Y.,   {Sijacki} D.,  2023, \mn@doi [arXiv e-prints]
  {10.48550/arXiv.2307.13733}, \href
  {https://ui.adsabs.harvard.edu/abs/2023arXiv230713733R} {p. arXiv:2307.13733}

\bibitem[\protect\citeauthoryear{{Rosdahl}, {Blaizot}, {Aubert}, {Stranex}  \&
  {Teyssier}}{{Rosdahl} et~al.}{2013}]{Rosdahl13}
{Rosdahl} J.,  {Blaizot} J.,  {Aubert} D.,  {Stranex} T.,   {Teyssier} R.,
  2013, \mn@doi [\mnras] {10.1093/mnras/stt1722}, \href
  {https://ui.adsabs.harvard.edu/abs/2013MNRAS.436.2188R} {436, 2188}

\bibitem[\protect\citeauthoryear{{Rosdahl}, {Schaye}, {Dubois}, {Kimm}  \&
  {Teyssier}}{{Rosdahl} et~al.}{2017}]{Rosdahl17}
{Rosdahl} J.,  {Schaye} J.,  {Dubois} Y.,  {Kimm} T.,   {Teyssier} R.,  2017,
  \mn@doi [\mnras] {10.1093/mnras/stw3034}, \href
  {https://ui.adsabs.harvard.edu/abs/2017MNRAS.466...11R} {466, 11}

\bibitem[\protect\citeauthoryear{{Rosdahl} et~al.,}{{Rosdahl}
  et~al.}{2022}]{Rosdahl22}
{Rosdahl} J.,  et~al., 2022, \mn@doi [\mnras] {10.1093/mnras/stac1942}, \href
  {https://ui.adsabs.harvard.edu/abs/2022MNRAS.515.2386R} {515, 2386}

\bibitem[\protect\citeauthoryear{{Rosen} \& {Bregman}}{{Rosen} \&
  {Bregman}}{1995}]{Rosen95}
{Rosen} A.,  {Bregman} J.~N.,  1995, \mn@doi [\apj] {10.1086/175303}, \href
  {https://ui.adsabs.harvard.edu/abs/1995ApJ...440..634R} {440, 634}

\bibitem[\protect\citeauthoryear{{Rudie}, {Steidel}, {Pettini}, {Trainor},
  {Strom}, {Hummels}, {Reddy}  \& {Shapley}}{{Rudie} et~al.}{2019}]{Rudie19}
{Rudie} G.~C.,  {Steidel} C.~C.,  {Pettini} M.,  {Trainor} R.~F.,  {Strom}
  A.~L.,  {Hummels} C.~B.,  {Reddy} N.~A.,   {Shapley} A.~E.,  2019, \mn@doi
  [\apj] {10.3847/1538-4357/ab4255}, \href
  {https://ui.adsabs.harvard.edu/abs/2019ApJ...885...61R} {885, 61}

\bibitem[\protect\citeauthoryear{{Salem}, {Bryan}  \& {Corlies}}{{Salem}
  et~al.}{2016}]{Salem16}
{Salem} M.,  {Bryan} G.~L.,   {Corlies} L.,  2016, \mn@doi [\mnras]
  {10.1093/mnras/stv2641}, \href
  {https://ui.adsabs.harvard.edu/abs/2016MNRAS.456..582S} {456, 582}

\bibitem[\protect\citeauthoryear{{Schaye} et~al.,}{{Schaye}
  et~al.}{2015}]{Schaye15}
{Schaye} J.,  et~al., 2015, \mn@doi [\mnras] {10.1093/mnras/stu2058}, \href
  {https://ui.adsabs.harvard.edu/abs/2015MNRAS.446..521S} {446, 521}

\bibitem[\protect\citeauthoryear{{Schroetter} et~al.,}{{Schroetter}
  et~al.}{2016}]{Schroetter16}
{Schroetter} I.,  et~al., 2016, \mn@doi [\apj] {10.3847/1538-4357/833/1/39},
  \href {https://ui.adsabs.harvard.edu/abs/2016ApJ...833...39S} {833, 39}

\bibitem[\protect\citeauthoryear{{Schroetter} et~al.,}{{Schroetter}
  et~al.}{2019}]{Schroetter19}
{Schroetter} I.,  et~al., 2019, \mn@doi [\mnras] {10.1093/mnras/stz2822}, \href
  {https://ui.adsabs.harvard.edu/abs/2019MNRAS.490.4368S} {490, 4368}

\bibitem[\protect\citeauthoryear{{Schroetter} et~al.,}{{Schroetter}
  et~al.}{2021}]{Schroetter21}
{Schroetter} I.,  et~al., 2021, \mn@doi [\mnras] {10.1093/mnras/stab1447},
  \href {https://ui.adsabs.harvard.edu/abs/2021MNRAS.506.1355S} {506, 1355}

\bibitem[\protect\citeauthoryear{{Semenov}, {Kravtsov}  \&
  {Caprioli}}{{Semenov} et~al.}{2021}]{Semenov21}
{Semenov} V.~A.,  {Kravtsov} A.~V.,   {Caprioli} D.,  2021, \mn@doi [\apj]
  {10.3847/1538-4357/abe2a6}, \href
  {https://ui.adsabs.harvard.edu/abs/2021ApJ...910..126S} {910, 126}

\bibitem[\protect\citeauthoryear{{Springel} et~al.,}{{Springel}
  et~al.}{2018}]{Springel18}
{Springel} V.,  et~al., 2018, \mn@doi [\mnras] {10.1093/mnras/stx3304}, \href
  {https://ui.adsabs.harvard.edu/abs/2018MNRAS.475..676S} {475, 676}

\bibitem[\protect\citeauthoryear{{Strong}, {Moskalenko}  \& {Ptuskin}}{{Strong}
  et~al.}{2007}]{Strong07}
{Strong} A.~W.,  {Moskalenko} I.~V.,   {Ptuskin} V.~S.,  2007, \mn@doi [Annual
  Review of Nuclear and Particle Science]
  {10.1146/annurev.nucl.57.090506.123011}, \href
  {https://ui.adsabs.harvard.edu/abs/2007ARNPS..57..285S} {57, 285}

\bibitem[\protect\citeauthoryear{{Suresh}, {Nelson}, {Genel}, {Rubin}  \&
  {Hernquist}}{{Suresh} et~al.}{2019}]{Suresh19}
{Suresh} J.,  {Nelson} D.,  {Genel} S.,  {Rubin} K. H.~R.,   {Hernquist} L.,
  2019, \mn@doi [\mnras] {10.1093/mnras/sty3402}, \href
  {https://ui.adsabs.harvard.edu/abs/2019MNRAS.483.4040S} {483, 4040}

\bibitem[\protect\citeauthoryear{{Tchernyshyov} et~al.,}{{Tchernyshyov}
  et~al.}{2023}]{Tchernyshyov23}
{Tchernyshyov} K.,  et~al., 2023, \mn@doi [\apj] {10.3847/1538-4357/acc86a},
  \href {https://ui.adsabs.harvard.edu/abs/2023ApJ...949...41T} {949, 41}

\bibitem[\protect\citeauthoryear{{Teyssier}}{{Teyssier}}{2002}]{Teyssier02}
{Teyssier} R.,  2002, \mn@doi [\aap] {10.1051/0004-6361:20011817}, \href
  {https://ui.adsabs.harvard.edu/abs/2002A&A...385..337T} {385, 337}

\bibitem[\protect\citeauthoryear{{Teyssier}, {Fromang}  \& {Dormy}}{{Teyssier}
  et~al.}{2006}]{Teyssier06}
{Teyssier} R.,  {Fromang} S.,   {Dormy} E.,  2006, \mn@doi [Journal of
  Computational Physics] {10.1016/j.jcp.2006.01.042}, \href
  {https://ui.adsabs.harvard.edu/abs/2006JCoPh.218...44T} {218, 44}

\bibitem[\protect\citeauthoryear{{Thomas} \& {Pfrommer}}{{Thomas} \&
  {Pfrommer}}{2019}]{Thomas19}
{Thomas} T.,  {Pfrommer} C.,  2019, \mn@doi [\mnras] {10.1093/mnras/stz263},
  \href {https://ui.adsabs.harvard.edu/abs/2019MNRAS.485.2977T} {485, 2977}

\bibitem[\protect\citeauthoryear{{Thomas}, {Pfrommer}  \& {Pakmor}}{{Thomas}
  et~al.}{2023}]{Thomas23}
{Thomas} T.,  {Pfrommer} C.,   {Pakmor} R.,  2023, \mn@doi [\mnras]
  {10.1093/mnras/stad472}, \href
  {https://ui.adsabs.harvard.edu/abs/2023MNRAS.521.3023T} {521, 3023}

\bibitem[\protect\citeauthoryear{{Toro}, {Spruce}  \& {Speares}}{{Toro}
  et~al.}{1994}]{Toro94}
{Toro} E.~F.,  {Spruce} M.,   {Speares} W.,  1994, \mn@doi [Shock Waves]
  {10.1007/BF01414629}, \href
  {https://ui.adsabs.harvard.edu/abs/1994ShWav...4...25T} {4, 25}

\bibitem[\protect\citeauthoryear{{Trebitsch}, {Blaizot}, {Rosdahl}, {Devriendt}
   \& {Slyz}}{{Trebitsch} et~al.}{2017}]{Trebitsch17}
{Trebitsch} M.,  {Blaizot} J.,  {Rosdahl} J.,  {Devriendt} J.,   {Slyz} A.,
  2017, \mn@doi [\mnras] {10.1093/mnras/stx1060}, \href
  {https://ui.adsabs.harvard.edu/abs/2017MNRAS.470..224T} {470, 224}

\bibitem[\protect\citeauthoryear{{Trotta}, {J{\'o}hannesson}, {Moskalenko},
  {Porter}, {Ruiz de Austri}  \& {Strong}}{{Trotta} et~al.}{2011}]{Trotta11}
{Trotta} R.,  {J{\'o}hannesson} G.,  {Moskalenko} I.~V.,  {Porter} T.~A.,
  {Ruiz de Austri} R.,   {Strong} A.~W.,  2011, \mn@doi [\apj]
  {10.1088/0004-637X/729/2/106}, \href
  {https://ui.adsabs.harvard.edu/abs/2011ApJ...729..106T} {729, 106}

\bibitem[\protect\citeauthoryear{{Tumlinson}, {Peeples}  \& {Werk}}{{Tumlinson}
  et~al.}{2017}]{Tumlinson17}
{Tumlinson} J.,  {Peeples} M.~S.,   {Werk} J.~K.,  2017, \mn@doi [\araa]
  {10.1146/annurev-astro-091916-055240}, \href
  {https://ui.adsabs.harvard.edu/abs/2017ARA&A..55..389T} {55, 389}

\bibitem[\protect\citeauthoryear{{Turner}, {Schaye}, {Steidel}, {Rudie}  \&
  {Strom}}{{Turner} et~al.}{2014}]{Turner14}
{Turner} M.~L.,  {Schaye} J.,  {Steidel} C.~C.,  {Rudie} G.~C.,   {Strom}
  A.~L.,  2014, \mn@doi [\mnras] {10.1093/mnras/stu1801}, \href
  {https://ui.adsabs.harvard.edu/abs/2014MNRAS.445..794T} {445, 794}

\bibitem[\protect\citeauthoryear{{Tweed}, {Devriendt}, {Blaizot}, {Colombi}  \&
  {Slyz}}{{Tweed} et~al.}{2009}]{Tweed09}
{Tweed} D.,  {Devriendt} J.,  {Blaizot} J.,  {Colombi} S.,   {Slyz} A.,  2009,
  \mn@doi [\aap] {10.1051/0004-6361/200911787}, \href
  {https://ui.adsabs.harvard.edu/abs/2009A&A...506..647T} {506, 647}

\bibitem[\protect\citeauthoryear{{{\"U}bler}, {Naab}, {Oser}, {Aumer}, {Sales}
  \& {White}}{{{\"U}bler} et~al.}{2014}]{Ubler14}
{{\"U}bler} H.,  {Naab} T.,  {Oser} L.,  {Aumer} M.,  {Sales} L.~V.,   {White}
  S. D.~M.,  2014, \mn@doi [\mnras] {10.1093/mnras/stu1275}, \href
  {https://ui.adsabs.harvard.edu/abs/2014MNRAS.443.2092U} {443, 2092}

\bibitem[\protect\citeauthoryear{\VAN{Voort}{van de}{van de}~Voort, {Springel},
  {Mandelker}, {van den Bosch}  \& {Pakmor}}{\VAN{Voort}{van de}{van de}~Voort
  et~al.}{2019}]{vandeVoort19}
\VAN{Voort}{van de}{van de}~Voort F.,  {Springel} V.,  {Mandelker} N.,  {van
  den Bosch} F.~C.,   {Pakmor} R.,  2019, \mn@doi [\mnras]
  {10.1093/mnrasl/sly190}, \href
  {https://ui.adsabs.harvard.edu/abs/2019MNRAS.482L..85V} {482, L85}

\bibitem[\protect\citeauthoryear{{Veilleux} et~al.,}{{Veilleux}
  et~al.}{2022}]{Veilleux22}
{Veilleux} S.,  et~al., 2022, \mn@doi [\apj] {10.3847/1538-4357/ac3cbb}, \href
  {https://ui.adsabs.harvard.edu/abs/2022ApJ...926...60V} {926, 60}

\bibitem[\protect\citeauthoryear{{Werk} et~al.,}{{Werk} et~al.}{2014}]{Werk14}
{Werk} J.~K.,  et~al., 2014, \mn@doi [\apj] {10.1088/0004-637X/792/1/8}, \href
  {https://ui.adsabs.harvard.edu/abs/2014ApJ...792....8W} {792, 8}

\bibitem[\protect\citeauthoryear{{Werk} et~al.,}{{Werk} et~al.}{2016}]{Werk16}
{Werk} J.~K.,  et~al., 2016, \mn@doi [\apj] {10.3847/1538-4357/833/1/54}, \href
  {https://ui.adsabs.harvard.edu/abs/2016ApJ...833...54W} {833, 54}

\bibitem[\protect\citeauthoryear{{Wiener}, {Pfrommer}  \& {Oh}}{{Wiener}
  et~al.}{2017}]{Wiener17}
{Wiener} J.,  {Pfrommer} C.,   {Oh} S.~P.,  2017, \mn@doi [\mnras]
  {10.1093/mnras/stx127}, \href
  {https://ui.adsabs.harvard.edu/abs/2017MNRAS.467..906W} {467, 906}

\bibitem[\protect\citeauthoryear{{Zabl} et~al.,}{{Zabl} et~al.}{2019}]{Zabl19}
{Zabl} J.,  et~al., 2019, \mn@doi [\mnras] {10.1093/mnras/stz392}, \href
  {https://ui.adsabs.harvard.edu/abs/2019MNRAS.485.1961Z} {485, 1961}

\bibitem[\protect\citeauthoryear{{Zabl} et~al.,}{{Zabl} et~al.}{2020}]{Zabl20}
{Zabl} J.,  et~al., 2020, \mn@doi [\mnras] {10.1093/mnras/stz3607}, \href
  {https://ui.adsabs.harvard.edu/abs/2020MNRAS.492.4576Z} {492, 4576}

\makeatother
\end{thebibliography}

\bsp	
\label{lastpage}
\end{document}